\documentclass[useAMS,usenatbib]{mn2e}

\usepackage{graphicx,amssymb,subfig,bm,xfrac}
\usepackage{float}
\usepackage[fleqn]{amsmath}  
\usepackage{color}

\newcommand{\be}{\begin{equation}}
\newcommand{\ee}{\end{equation}}
\newcommand{\ba}{\begin{eqnarray}}
\newcommand{\ea}{\end{eqnarray}}

\def\simlt{\lower.5ex\hbox{$\; \buildrel < \over \sim \;$}}
\def\simgt{\lower.5ex\hbox{$\; \buildrel > \over \sim \;$}}

\usepackage{graphicx} 

\title[Dark Matter Astrometry]{Dark Matter Astrometry: Accuracy of sub-halo positions for the measurement of self-interaction cross sections}
\author[D.\ Harvey et al.]{David Harvey$^{1}$\thanks{e-mail: {\tt drh@roe.ac.uk}}, 
Richard Massey$^{2,1}$, Thomas Kitching$^{1}$, Andy Taylor$^{1}$,
Eric Jullo$^{3}$, \newauthor Jean-Paul Kneib$^{3}$, Eric Tittley$^{1}$ \& Philip J. Marshall$^{4}$ \\
$^{1}$ SUPA, University of Edinburgh, Royal Observatory, Blackford Hill, Edinburgh EH9 3HJ, UK  \\ 
$^{2}$ Institute for Computational Cosmology, Durham University, South Road, Durham DH1 3LE, UK  \\
$^{3}$ Laboratoire d'Astrophysique de Marseille, CNRS-Universit\'e Aix-Marseille, 38 rue F.\ Joliot-Curie, 13388 Marseille, France\\
$^{4}$ Department of Physics, University of Oxford, Keble Road, Oxford, OX1 3RH, UK}
\begin{document}

\date{Accepted ---. Received ---; in original form \today.}

\pagerange{\pageref{firstpage}--\pageref{lastpage}} \pubyear{2011}

\maketitle

\label{firstpage}

\begin{abstract}


 
\noindent Direct evidence for the existence of dark matter and measurements of its interaction cross-section have been provided by the physical offset between dark matter and intra-cluster gas in merging systems like the Bullet Cluster.
%
Although a smaller signal, this effect is more abundant in minor mergers where infalling substructure dark matter and gas are segregated.
In such low-mass systems the gravitational lensing signal comes primarily from weak lensing.
A fundamental step in determining such an offset in substructure is the ability to accurately measure the positions of dark matter sub-peaks.
Using simulated Hubble Space Telescope observations, we make a first assessment of the precision and accuracy with which we can measure infalling groups using weak gravitational lensing.
We demonstrate that using an existing and well-used mass reconstruction algorithm can measure the
positions of $1.5\times10^{13}M_\odot$ substructures that have parent halos ten times more massive with a bias of less than $0.3\arcsec$.
In this regime, our analysis suggests the precision is sufficient to detect  (at 3 $\sigma$ statistical significance) the expected mean offset between dark matter and baryonic gas in infalling groups from a sample of $\sim50$ massive clusters. 

\end{abstract}


\section{Introduction} \label{sec:intro}

Evidence for dark matter (DM) has been accumulating for 80 years \citep[e.g.][]{1933AcHPh...6..110Z,1980ApJ...238..471R,2000FCPh...21....1B, 2006ApJ...648L.109C}, yet its nature and properties remain poorly understood. 
How DM manifests itself in the Universe has become a key question in both particle and astrophysics, which has resulted in a variety of studies all attempting to shed some light on this dark mystery. 

Despite evidence from accumulating from astronomical sources, the tightest constraints on the properties of DM are being led by terrestrial experiments. The Large Hadron Collider is attempting to create new particles in high energy proton-proton collisions that could potentially be dark matter candidates \citep[e.g.][]{2011JPhCS.335a2003M, 2009patl.book..179B}. 
Furthermore, Direct Detection experiments are trying to observe the galactic dark matter wind caused by the orbit of the solar system around the Galaxy, and the Earth around the Sun \citep[e.g.][]{2010EPJC...67...39B, 2011arXiv1109.0702A, 2005PhR...405..279B, 2009APh....31..261B}. 

Astronomical techniques currently provide looser constraints on otherwise unaccessible parameters. 
DM annihilation signals could be observed through gamma rays originating from the centre of the galaxy \citep[e.g][]{2011PhLB..697..412H, 2009PhRvD..80l3511C}, placing constraints on the annihilation cross section. Alternatively the density profiles of galaxy clusters can constrain the self interaction cross section. \citet{2000ApJ...544L..87Y} found that a few collisions per particle per Hubble time can significantly affect the profile at the core of the cluster. Moreover, cosmological simulations using self interacting dark matter have shown that a small but finite cross section will have an affect on the core size and central density \citep{2012arXiv1208.3025R} . \citet{2012arXiv1208.3026P} tried to compare these simulations with observations and found that constraints from  such a technique will most probably be improved by measurements of central densities and not halo shapes. 
Both techniques provide a unique way to probe the properties of DM.

The trajectories of different mass components during major mergers like the Bullet Cluster (1E 0657-558) have recently provided important constraints on the DM--DM self-interaction cross-section $\sigma$ \citep{2004ApJ...604..596C,2006ApJ...652..937B,2006ApJ...648L.109C, 2007ApJ...668..806M}.
%
The baryonic components of a galaxy cluster can be seen via direct imaging: optical emission from galaxies, and X-ray bremsstrahlung emission from hot gas. Dark matter cannot be observed directly, but its projected density can be reconstructed from the observable ``gravitational lensing'' of the images of background sources behind the cluster \citep[see reviews][]{2001PhR...340..291B,2003ARA&A..41..645R, 2008ARNPS..58...99H,2010RPPh...73h6901M}. 
In the model for the Bullet Cluster collision, the
X-ray emitting gas in the intracluster medium was slowed during its first core passage, preventing it from travelling far from the point of impact.
The dark matter in each cluster, interacting essentially only via gravity and mapped using its weak and strong gravitational lensing effects, is supposed to have passed through unaffected.
In this picture,
the temporary separation of the Bullet Cluster's DM from its gas yields a constraint of $\sigma / m < 1{\rm cm^2g^{-1}}$ \citep{2004ApJ...606..819M}. 
Similar analyses yield $\sigma / m < 4{\rm cm^2g^{-1}}$ from cluster MACSJ0025-1222 \citep{2008ApJ...687..959B}
and $\sigma / m < 3{\rm cm^2g^{-1}}$ from cluster Abell 2744 \citep{2011MNRAS.417..333M}.


As shown by \citet{2011MNRAS.415..448W} a further displacement can be measured between the DM and the stars of the cluster.
Assuming that stars act as non-interacting tracer particles, any collision-induced separation between the DM density peaks (inferred from their lensing effects) and the cluster member galaxies (visible by their starlight) could provide evidence for there having been some self-interacting dark matter.   
The offset between DM and stars in cluster Abell 3827 intriguingly suggests a non-zero interaction cross-section, with {\it lower} limit $\sigma / m > 4.5\times10^{-7}{\rm (t/10^{10}yr)^{-2}cm^{2}g^{-1}}$, where $t$ is the infall time of the sub clump around the main halo \citep{2011MNRAS.415..448W}.
This result is, however, sensitive to the interpretation of a very small number of proposed multiple (strongly-lensed) images.


In this work we assess the precision and accuracy of weak gravitational lensing measurements of the {\it position} of mass peaks.
This differs from the many studies that have assessed the precision and accuracy of measurement of the {\it mass} of mass peaks.
We investigate whether it will be possible to detect small offsets in position on the sky between the baryonic and DM density peaks of cluster substructures \citep[][hereafter MKN]{2011MNRAS.413.1709M}.
We imagine detecting these infalling galaxy groups, and measuring their barycenters, from their X-ray (or optical) emission, and comparing with the positions of mass density peaks reconstructed by analysis of weakly lensed background objects in the vicinity. 
Such analyses have been carried out in individual interacting clusters using flexible exploratory mapping techniques by \citep[e.g.\ ][]{2004ApJ...604..596C, 2004ApJ...606..819M}; here we consider measuring offsets -- ``bulleticities'' -- in many different clusters, and combining the results in a statistical measurement of the interaction cross-section (MKN).
In particular, we are interested in using analytically simulated data to answer the following questions: 
\begin{itemize}
\item To what precision can we measure the offset in a single infalling substructure? 
\item Can we identify a point estimator whose simple combination over a sample will provide a measurement of sub-halo position with minimal bias?
\item What are the dominant sources of residual bias in this estimate?
\item How large a sample of observed clusters are we likely to need to be able to detect an offset between dark matter and baryonic as predicted by MKN?
\item What further investigation might be needed to prove the utility of this technique for probing DM interaction cross-sections?
\end{itemize}


This paper is organised as follows. In Section~\ref{sec:theory} we outline the theory behind weak gravitational lensing and our goals in its application.
In Section~\ref{sec:method} we present an end-to-end simulation pipeline in which we start with a known mass distribution, simulate HST lensing data, then use {\tt Lenstool} to reconstruct the mass distribution. 
In Section~\ref{sec:results} we describe our results. 
In Section ~\ref{sec:disc} and ~\ref{sec:conc} we conclude and outline future work.

\section{Theory}\label{sec:theory}

\subsection{Weak Gravitational Lensing}

Gravitational lensing is the deflection of light rays by the the distortion of space-time around any massive object.
This phenomenon can be used to map the distribution of mass, including otherwise invisible dark matter.


A 3D density distribution $\rho$ with gravitational potential $\Phi$ can be projected onto the plane of the sky to obtain a 2D deflection potential
\begin{equation}
\Psi ~~\equiv~~ \frac{D_{\rm OL} D_{\rm LS}}{D_{\rm OS}} \frac{2}{c^2}\int \Phi~\mathrm{d} z, \label{eqn:psi}
\end{equation}
where the angular diameter distances between the Observer, Lens and Source encode the geometry of a convergent lens.
The image of a background galaxy passing through this potential is magnified by a convergence
\begin{equation}
\kappa ~~=~~ \frac{1}{2} \left(\frac{\partial^2\Psi}{\partial x^2} + \frac{\partial^2\Psi}{\partial y^2}\right)\label{eqn:convergencedefn}
\end{equation}
and distorted by a shear 
\begin{eqnarray}
\gamma_1 &=& \frac{1}{2} \left(\frac{\partial^2\Psi}{\partial x^2} - \frac{\partial^2\Psi}{\partial y^2}\right)  \\
\gamma_2 &=& \frac{\partial^2\Psi}{\partial x\partial y} ~,
\label{eqn:sheardefn}
\end{eqnarray}
where $\gamma_1 (\gamma_2) $ refers to elongation along (at $45^\circ$ to) an $x$~axis defined arbitrarily in the plane of the sky \citep{2001PhR...340..291B,2003ARA&A..41..645R}. 

Around a foreground galaxy cluster, background galaxies appear aligned in distinctive circular patterns, with tangential and curl components
\begin{eqnarray}
\label{eqn:tangential}
\gamma_t&=&-\left[\gamma_1\cos(2\phi)+\gamma_2\sin(2\phi)\right] \\
\label{eqn:cross}
\gamma_\times&=&-\gamma_1\sin(2\phi)+\gamma_2\cos(2\phi) 
\end{eqnarray}
where $\phi$ is the angle of the galaxy position with respect to the a Cartesian axis. 

If galaxies were intrinsically circular and of fixed size, the applied shear would simply change their apparent ellipticity.
In practice it is necessary to average the observed shapes of $\sim100$ galaxies to remove the influence of their complex morphologies;
and because of a degeneracy between shear and magnification, only the `reduced shear'
\begin{equation}
g\equiv\frac{\gamma}{1-\kappa}
\end{equation}
is observable.

\subsection{Bulleticity}\label{sec:bullet}

Attempts to constrain the self interaction cross section of dark matter from major galaxy cluster collisions face two obstacles.
First, measuring a separation between the dark matter and baryonic gas requires a merger between clusters of similar masses to be seen at just the right time since first core passage, and these are rare events \citep{2010MNRAS.408.1277S}.
Second, uncertainties in the impact velocity, impact parameter and orientation with respect to the line of sight severely limit constraints from individual clusters \citep{2009ApJ...700.1404R}.

Fortunately, hydrodynamical simulations of structure formation predict a mean offset between DM and gas during the infall of {\it all} sub-halos into massive clusters (MKN).
Although weak gravitational lensing cannot precisely resolve the positions and masses of individual pieces of small substructure, a statistical ``bulleticity'' signal can be obtained by averaging the measurements from many clusters.
The bulleticity vector $\bm{b}$ is the offset between substructure's total mass (where dark matter dominates) and baryonic components in the plane of the sky
\begin{equation}
\bm{b}\equiv b_r\hat{\bm{e}}_r+b_t\hat{\bm{e}}_t,
\label{bullet}
\end{equation}
where $\hat{\bm{e}}_r$ and $\hat{\bm{e}}_t$ are unit vectors in the radial and tangential directions with respect to the cluster centre. 
Hydrodynamical simulations show that, despite complex and interacting processes, the net effect of cluster gastrophysics is a force on the substructure gas similar to a simple buoyancy that produces an offset $\langle b_r\rangle>0$.
This is the key signal in which we are interested.
The simulations also show that, with a sufficiently large sample and no preferred in-fall handedness, $\langle b_t\rangle\equiv0$.
Checking that measurements of this are consistent with zero will be a useful test for residual systematics.


MKN showed that for a $\Lambda$CDM paradigm with collisionless dark matter, the expected radial offset between baryonic and dark components of substructure is $\sim10\arcsec,3.5\arcsec$ and $2\arcsec$ at a redshift $z=0.1,0.3,0.6$ respectively at a radial of distance of $0.15r_{500}$, which increased towards to the centre of the cluster.
Therefore the measurement of an offset relies upon an ability to measure the position of substructure components with minimal bias near to the core. 
Statistical errors will be gradually beaten down by averaging many measured offsets. 
However any systematic bias in the centroid of either component will propagate into constraints on the interaction cross-section.


The spatial resolution of the X-ray space telescope Chandra is sub-arcsecond, whereas any weak lensing map will be limited by the finite density of resolved galaxies to $\sim10''$ with the deepest, highest resolution data \citep[e.g.][]{2007Natur.445..286M} .
Although the accuracy in which one can define the X-ray peak will depend on the distribution of gas the dark matter peak will be affected by a similar problem. 
Therefore it is possible to assume that the error in the X-ray peak position is subdominant to that of dark matter, and study in detail the reliability of weak lensing centroid estimates only.
We will also quantify the precision of weak lensing centroiding, to estimate the sample size required to detect bulleticity.

\section{Method}\label{sec:method}

\subsection{Simulated Shear Fields}\label{sec:sims}

In order to examine the exact behaviour of weak lensing as a positional estimate of dark matter we need initial experiments in carefully controlled environments.
This includes having mass distributions with well defined correct answers rather than cosmological simulations.
We therefore create simulated shear fields containing DM halos of known position, mass and ellipticity.
For an analytic model, 

we adopt the \citet[][NFW]{1996ApJ...462..563N} density profile for a galaxy cluster at a conservative redshift 0f 0.6,
\begin{equation}
\label{eq:NFW}
\rho(r)\propto\frac{1}{\frac{r}{r_s}\left(1+\frac{r}{r_s}\right)^2}
\end{equation}
where  the scale radius $r_{\rm s}$ can be expressed in terms of the concentration parameter $c=\sfrac{r_{\rm vir}}{r_{\rm s}} $ (and $r_{\rm vir}$ is the virial radius). 
For typical clusters, an empirical relation \citep{2008MNRAS.391.1940M} suggests

\begin{equation}
\log \langle c\rangle=0.830 - 0.098\log\left(M_{\rm vir}/\left[10^{12}h^{-1}M_\odot\right]\right),
\label{eqn:conc}
\end{equation}

\noindent where $M_{\rm vir}$ is the virial mass.

\begin{figure}
 \includegraphics[width = 8.5cm]{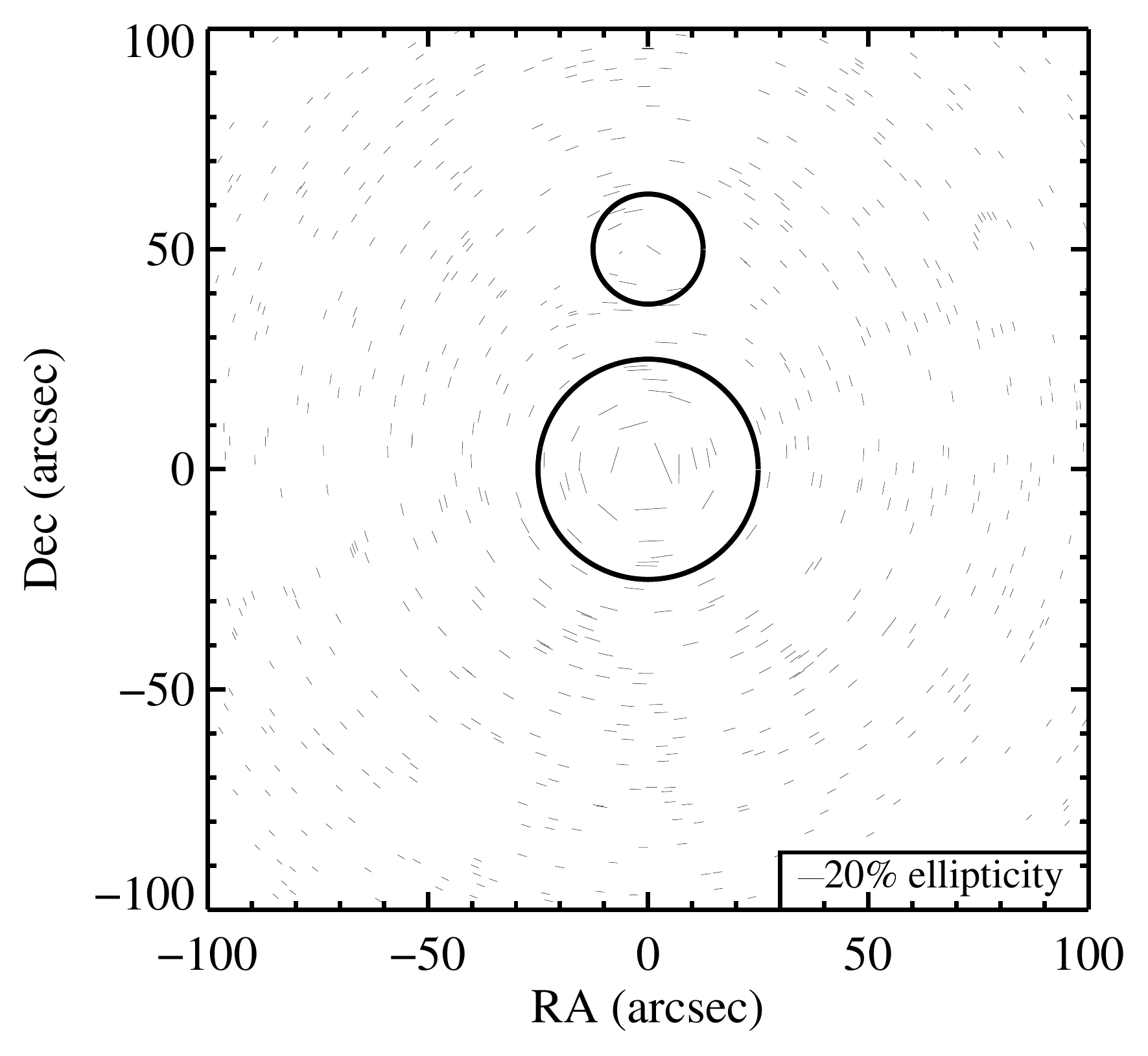}
 \caption{\label{fig:simulation}
 The reduced shear signal  of a simulated cluster with a NFW profile. The main halo has a $M_{200}$ of $8\times10^{14} M_\odot$ and is positioned at (0,0) and the sub-halo, $8\times10^{13}M_\odot$, is positioned at (0,49). The field of view represents that of a Hubble Space telescope Advanced Camera for Surveys with a typical density of galaxies of 80/square arc minute. The circles are a guide for where they are placed and have no physical significance.}
\end{figure}

Using multiple NFW halos, we construct a cluster system with infalling galaxy group(s).
As shown in Figure~\ref{fig:simulation}, the baseline configuration includes a main halo in the centre of the field of view plus a sub-halo $49\arcsec$ to the north.
Substructure typically contains $\sim10\%$ of the mass of a system \citep[e.g.][]{2012MNRAS.419.1017C}, 
we fix the mass of the sub-halo to be always 10\% that of the parent halo (with a concentration parameter given by equation~\ref{eqn:conc}).

The shears due to multiple mass components in the same cluster simply add, such that
$\gamma_{\rm total}=\gamma_{\rm main~halo}+\gamma_{\rm sub-halo_1}+\gamma_{\rm sub-halo_2}+\dots+\gamma_{\rm sub-halo_n}$.

In order to get high signal to noise imaging of cluster cores we manufacture simulated Hubble Space Telescope (HST) weak lensing measurements of the cluster system.
In a $200\arcsec\times200\arcsec$ field of view, shear measurements are simulated from 80 (randomly placed) galaxies per square arc minute, as could be obtained from a full 1-orbit exposure in the F814W band with the Advanced Camera for Surveys (ACS). In a typical Hubble archive the redshift of a cluster is varies between 0 and 1. We choose a conservative redshift of 0.6, at which the bulleticity signal is small but potentially detectable. 
For each cluster configuration and mass, we generate 100 noise realisations.

\subsubsection{Intrinsic Galaxy Morphologies}

The basic challenge with weak lensing measurements is that galaxies are not inherently circular -- indeed, the ellipticity of a typical galaxy is an order of magnitude larger than the shear. 
Chance alignments of galaxies can mimic a coherent gravitational lensing signal and introduce noise into the mass reconstruction.
In individual clusters, such noise can cause centroid shifts of $\sim10\arcsec$, but this should average away as long as the noise has no preferred direction over a sample of clusters.
Leauthaud et al. (2007) state that galaxies typically have a intrinsic ellipticity distribution with a mean of zero and an width, $e_{\rm int}$ of $0.3$.  This can be expressed as a complex number by $e_{\rm int} = |e_{\rm int}|\exp(2i\theta)$, where $\theta$, is the angle of the galaxy. One can then transform the galaxy from the source plane to the image plane using the complex reduced shear, $g=|g|\exp(2i\phi)$, where $\phi$ is orientation of the galaxy due to the lensing affect, via,
\begin{equation}
e^{\rm (I)}=\frac{ e_{\rm int}  +2g +g^2e_{\rm int}^* }{1+|g^2|+2\Re(ge_{\rm int}^*)},
\label{eqn:e_image}
\end{equation}
where the star represents the complex conjugate.

\subsubsection{Elliptical Mass Distributions}

Galaxy clusters are 
often not 
spherically symmetric \citep{2002astro.ph..1421J}.
Misidentifying the shape of a halo can introduce spurious detections of substructure along the major axis, or shift the apparent position of real substructure. 
It is therefore important to check whether elliptical halos affect the centroid estimate of both the cluster and the sub-halo. 
We have run simulations with both a spherical and an elliptical main halo.
In elliptical cases, the ellipticity of the main potential was fixed at 0.2 (where ${\rm ellipticity}=[a^2-b^2]/[a^2+b^2]$). To span a range of possible scenarios, the major axis is aligned at $0^\circ$, $45^\circ$ or $90^\circ$ from the positive $x$-axis (in the latter case, this points towards the sub-halo). 

It was considered that force fitting a circularly symmetric fit to an elliptical  main halo could potentially bias the position however the signal to noise of the sub-halo would mean that constraining the ellipticity would not be possible.
Moreover DM would interact only gravitationally, and therefore we expect any infalling halo to retain its radial symmetry, thus in all cases the sub-halo is kept circular.


\begin{table*}
\begin{minipage}{126mm}
\begin{center}
  \caption{Input values and priors used during the reconstruction on the main and sub-halo in the simulations. The values in the square brackets refer to the range, and the dots refer to the different mass scales of the simulations. }
  \begin{tabular}{| c | c | c | c |}
    \hline 
     & Input Value & Prior & Type \\ \hline
    Main Position (arcseconds) & (0,0) &  $30$ on (0,0) & Gaussian \\
    Sub 1 Position (arcseconds) & (0,49) & $25\arcsec$ on (0,49) & Flat Circle  \\
    Sub 2 Position (arcseconds) & (49,49),(49,0),(0,-49) & $25\arcsec$ Radius & Flat Circle \\
    Main Halo Mass  ($M_\odot$) & $ (1,1.5 \dots 7.5,8)\times10^{14} $ & $ [0.5,49] \times 10^{14} $ & Flat \\
    Sub Halo Mass ($M_\odot$) & $ (1,1.5 \dots 7.5,8) \times10^{13} $ & $ [0.5,49] \times 10^{13} $ &  Flat \\
    Main Halo Concentration & Mass:Conc Rel & $[1,10]$ & Flat \\
    Sub Halo Concentration & Mass:Conc Rel & $[1,10]$ & Flat \\
    Mass Priors & $M_{main}>M_{sub} $& $M_{main}>M_{sub} $& Statement  \\ \hline
  \end{tabular}
  \label{tab:priors}
\end{center}
\end{minipage}
\end{table*}

\subsubsection{Imperfect Shape Measurement}\label{sec:shape}

Achieving sub-percent accuracy in the measurement of galaxies' apparent shapes is an ongoing challenge \citet{2006MNRAS.368.1323H, 2007MNRAS.376...13M, 2010MNRAS.405.2044B,G10results,MDM}.
Even with a {\it space}-based telescope, the point spread function (PSF) varies across the field of view and can change over time \citep{2007ApJS..172..203R}.
If the PSF is not accurately modelled or the image effectively deconvolved, it can be spuriously imprinted upon the shear measurements.
Image noise and pixelation further impede the measurement of small, faint galaxy shapes.

We do not consider multiplicative shear measurement biases here, since they bias only the recovered mass estimates, and not the positions.
We do, however, consider additive shear measurement biases, which will affect the inferred mass clump positions.
The PSF normally has a preferred direction with respect to the telescope, but the location of substructure and the angle of orientation at which the cluster is imaged will vary from cluster to cluster.
In each realisation of a simulated catalogue, we add a constant spurious signal $c_i$ to each component of shear, drawn from a Gaussian distribution with mean $0$ and standard deviation $0.01$ (which is split into shear components so that it is in a random direction).
Finally, we model the pixelation noise by adding an additional stochastic component to each shear measurement, drawn from a Gaussian distribution with mean $0$ and width $0.01$ again split into components, although this is effectively degenerate with (and subdominant to) the intrinsic ellipticity. Thus the observed shears become
\begin{equation}
e^{\rm tot}_i=e_i^{\rm (I)}+c_i+ \sigma_i^N.
\label{eqn:shape}
\end{equation}
Although many algorithms linearise the lensing potential, which doesn't hold in regimes of $g>0.1$, the bias introduced by this assumption can be considered a multiplicative factor to the shear and in fact would not have an affect to the position of the sub-halo \citep{2012MNRAS.tmp.3383M}

\subsubsection{Galaxy Redshift Distribution}\label{sec:rshift}

A statistical measurement of bulleticity will require a large sample of galaxy clusters, and multicolour imaging may not be available for them all.
The distortion experienced by each galaxy image depends upon the lensing geometry as described by equation~\eqref{eqn:psi}.
With only monochromatic imaging, it can even be impossible to tell whether galaxies are behind a cluster (and therefore lensed) or in front of it (and therefore undistorted). 
Allowing foreground galaxies in the galaxy catalogue will dilute the inferred shear signal.
We introduce a source galaxy redshift distribution
\begin{equation}
p(z)\propto z^2\exp\left[-\bigg(\frac{z}{z^\star}\bigg)^{1.5}\right],
\label{eqn:nz}
\end{equation}
where $z^\star=z_{\rm med}/1.1412$ and $z_{\rm med}=1.0$ \citep{2007MNRAS.374.1377T}.
We apply this uniformly across the field of view.
We assume we know exactly the redshift of the cluster (0.6) and for the purposes of measuring the position we do not concern ourself with $\Sigma_{\rm crit}$ and the total mass.

A further problem with having only monochromatic imaging is that it will be impossible to distinguish between background sources and cluster members.
Although the affect of this would be further dilution of the signal, since the members will be correlated with the density profile of the cluster the dilution will also be correlated.
It is therefore possible for the position of the halo to be biased if these galaxies are included in the reconstruction.
Therefore, a simple distribution of member galaxies is placed over the cluster such that they follow the NFW profile. 
The number of member galaxies is then increased and to study the affect the member galaxies may have.

\subsubsection{Multiple Substructures}

In the paradigm of hierarchical structure formation, clusters grow through multiple mergers, so multiple sub-halos may be physically close to a cluster at a given time.
The presence of multiple sub-halos
will complicate the shear field and thus make it harder to estimate the positions of each. 
It is therefore important to be confident that if sub-halos are close together in real space their signals do not cause a bias in any direction.
A set of realisations were run with a second sub-halo was introduced into the field.
The first sub-halo remained at (0,49) arcseconds from the main halo; to span a range of possible configurations, the second halo was placed at three different positions (49,49), (-49,0) and (0,-49) arcseconds from the main halo.

\subsubsection{Potential Line of Sight Contamination}

Independent large scale structure at different redshifts may happen to lie along the line of sight to the cluster, and be misinterpreted as substructure
\citep{2001A&A...370..743H,2012MNRAS.420.1384S}.
As unassociated galaxy groups will not be falling into the cluster, they will not exhibit any systematic offset between DM and gas, and their inclusion will spuriously dilute the measured bulleticity. 
Since substructure will be initially identified via X-ray imaging, we can estimate the number of coincidentally aligned structures by considering the density of X-ray luminous groups in unpointed observations.


In the 1.64 square degree COSMOS survey \citep{2007ApJS..172....1S}, \citet{2007ApJS..172..182F} found 206 X-ray groups with masses $10^{13}$--$10^{14} M_{\odot} h^{-1}_{72}$, which matches the mass range considered in this work.
\citet{2010ApJ...709...97L} tried to detect all of these groups via weak lensing from 1-orbit HST imaging.
About a quarter of the groups are detected at greater than our $2\sigma$ detection threshold.
Scaling this down to the $200\arcsec\times200\arcsec$ ACS field of view, we expect a contaminant of around 1 spurious peak for every 20 clusters.
This $\sim5\%$ dilution should be considered in a second larger survey (e.g. HSC, DES, Euclid), and could be reduced if (even coarse) photometric redshifts were available for some galaxies.

\subsubsection{Substructure as a function of distance from the cluster centre and mass ratio}
The basis simulation is set up such that the distance the sub-halo is from the centre of the cluster and the mass fraction between the main halo and the sub-halo is held constant.
Although the values used for the simulations are that of a typical cluster these will not be constant in the case of real data and therefore the radial distance and the mass fraction are both independently varied.




\begin{figure}
		\begin{centering}
	 		\subfloat[Main halo positional estimates]{\label{fig:asymm_1}\includegraphics[width = 7.5cm]{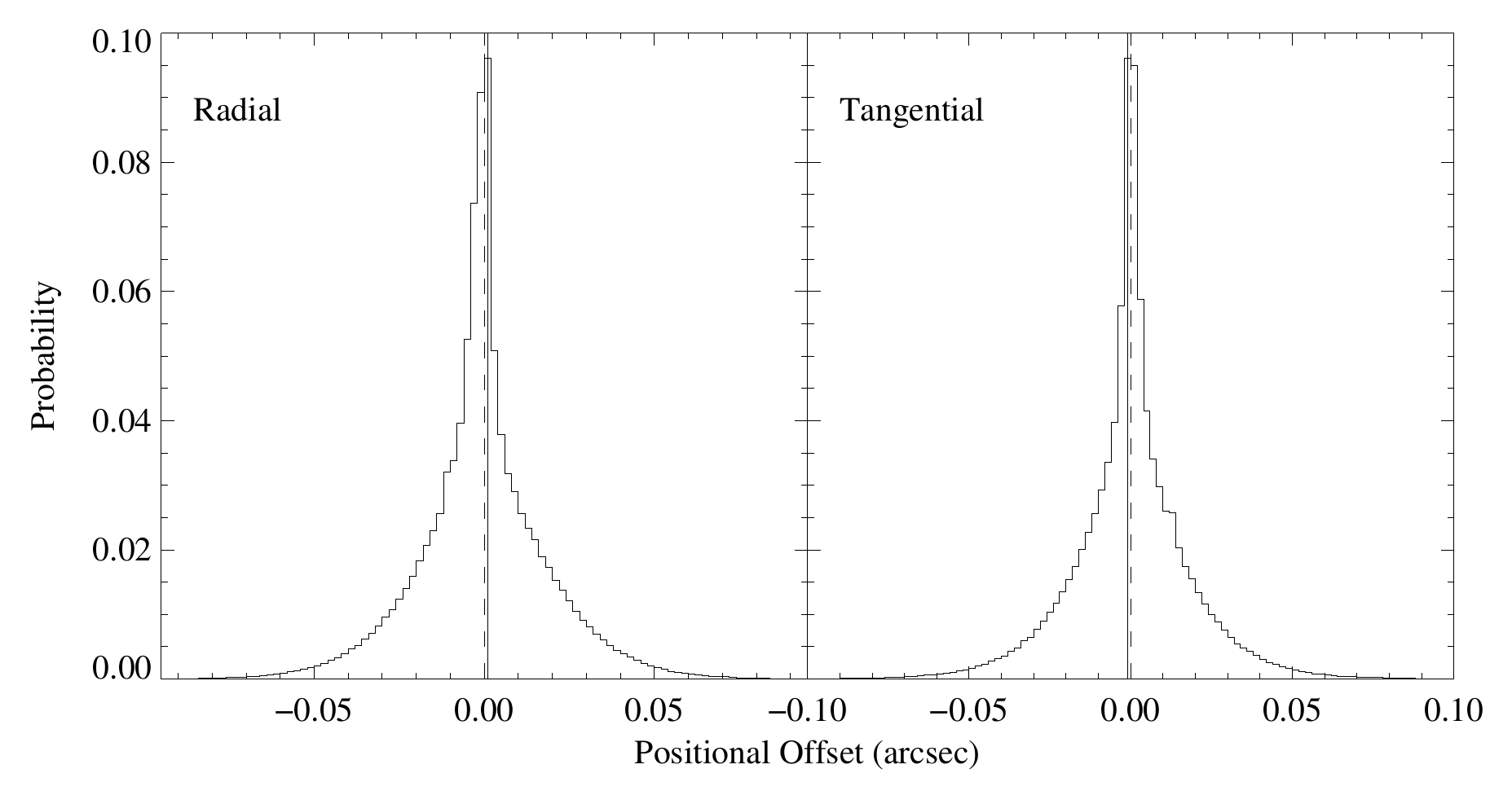}}\\
  			\subfloat[Sub-halo positional estimates]{\label{fig:asymm_2}\includegraphics[width = 7.5cm]{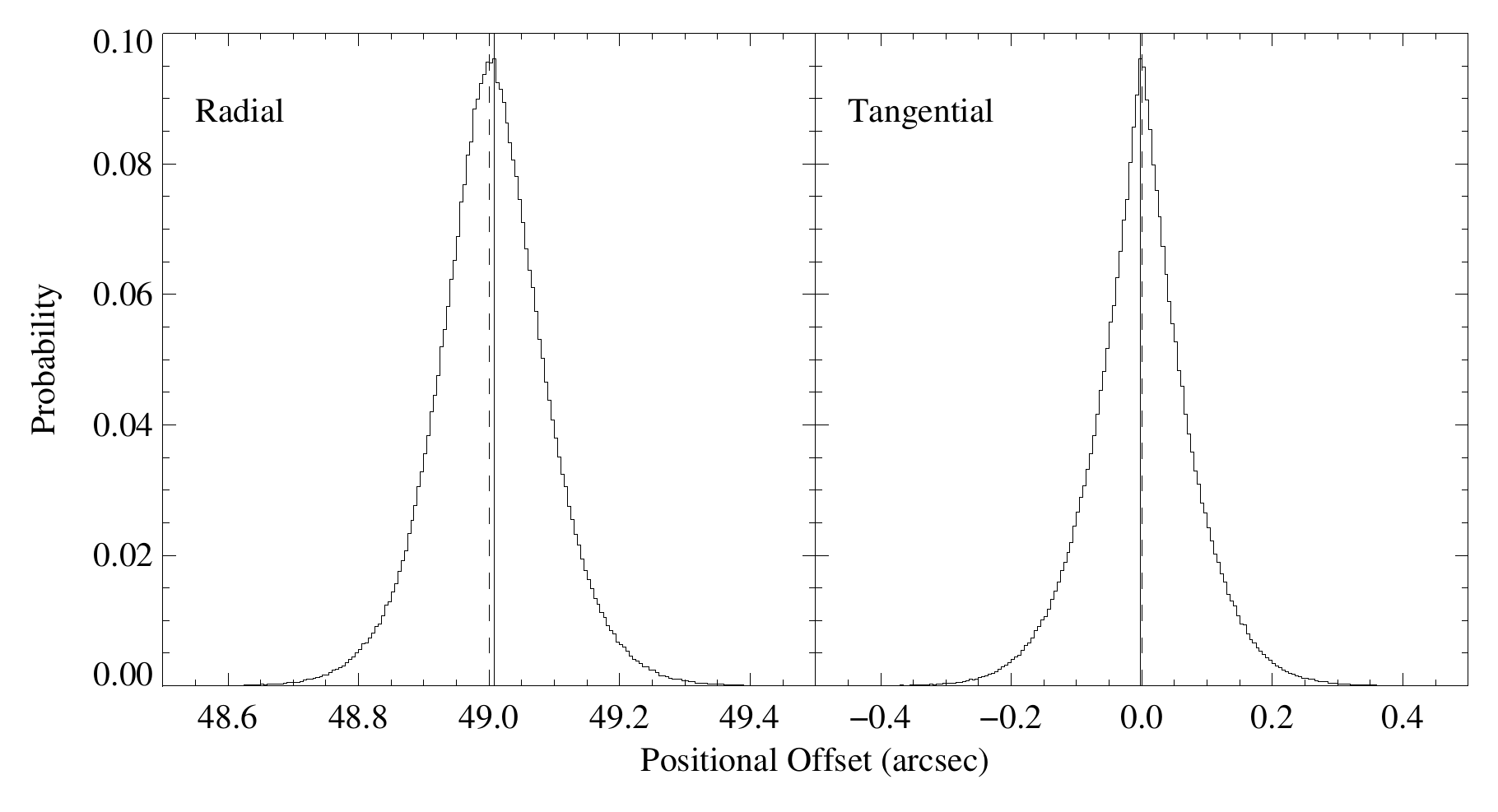}} 
		\caption{\label{fig:asymm}
			The likelihood surface for the main (top) and sub (bottom) halo positions in the case of zero noise (gravitational shear only). The binned histograms show the true posteriors and their maximum likelihoods as the solid line. The dotted line is the true position. The left hand panels are the position in the radial direction and the right hand panels are tangential direction. In the case of no noise the likelihood surface derived from {\tt Lenstool} exhibits no  bias around the maximum likelihood. 
			}
		\end{centering}

\end{figure}

\subsection{Mass Reconstruction}\label{sec:recon}

Many algorithms have been developed to reconstruct the mass, concentration and position of massive ($>10^{14}M_\odot$) halos from observations of weak (and strong) gravitational lensing \citep{2005A&A...437...39B,2006A&A...458..349C,2007MNRAS.375..958D,2009A&A...500..681M}.
However, testing of these has generally focussed on the mass and concentration parameters, positional accuracy has not yet been pushed to the low $\sim10^{13}M_\odot$ mass regime.


To determine the viability of bulleticity measurements, weak lensing reconstructions must be tested in scenarios that reflect the environments in which it will be used. 
The main requirement for bulleticity is an accurate estimate of the sub-halo and main halo positions with minimal bias.


{\tt Lenstool} (\citealt{2007NJPh....9..447J}) is open source software that calculates analytical models of the lensing signal for specific cluster density profiles and then compares them against the observed data. 
In a Bayesian framework, {\tt Lenstool} samples the posterior using a Markov Chain Monte Carlo algorithm.
It continually probes the entire parameter space providing an estimate of entire posterior surface. 
The posterior for a given prior, $\Pi$, and likelihood, $L$, is
\begin{equation}
P= L\Pi
\label{eqn:bayes}
\end{equation}
where the likelihood is Gaussian,
\begin{equation}
L=\frac{1}{(2\pi\sigma^2)^{N/2}}e^{-\chi^2/2}.
\end{equation}
The $\chi^2$ statistic in  {\tt Lenstool}  is calculated by converting the observed galaxy ellipticity to the source plane using the proposed model parameters. 
The resulting ellipticity should represent the intrinsic shape of the galaxy, which when summed over the entire field will have a mean of zero with some known variance.
A chi-squared close to one shows that the model parameters to convert the image to the source plane were a good fit. Thus the chi-squared for given set of parameters, calculated in the source plane, is simply
\begin{equation}
\chi^2=\sum^{N}_{i=1}\sum^{2}_{j=1}\frac{(e^{(s)}_{i,j})^2}{\sigma_{i,j}^2},
\label{eqn:bayesian}
\end{equation}

\noindent where the total error in the ellipticity, $\sigma$,  is the sum of the intrinsic ellipticity and the shape measurement error added in quadrature, i.e. $\sigma = \sqrt{\sigma_{int}^2+\sigma_{shape}^2}$.

The fact that the chi-squared is calculated in the source plane means the reconstruction can be affected by the way {\tt Lenstool} converts from the source plane to the image plane. 
The input parameters for {\tt Lenstool} are the semi major axis $a$, the semi minor axis, $b$, and the angle of the galaxy with respects to the image $x$-axis. These ellipse descriptors not only define the ellipticity of the galaxy but also the size.
It has been shown in \citet{2012ApJ...744L..22S}  that measuring the sizes of a galaxies is difficult and also ambiguous in how one defines it  therefore we would like to avoid using this parameter. 

We therefore decide to use \textit{option 7} in {\tt Lenstool}$^1$,
\footnotetext[1]{We also tested \textit{option 6} which takes $a$,$b$ and the angle and transforms them directly to the source plane via
\begin{equation}
Q ^{(s)}=AQA,
\label{eqn:matop}
\end{equation}
where Q is the quadrupole moment matrix, 
\begin{equation}
Q_{ij}=\frac{ \int d^2\theta q_I[ I(\overrightarrow{\theta}) ](\theta_i-\Theta_i)(\theta_j-\theta_j)} {\int d^2\theta q_I[ I(\overrightarrow{\theta}) ]}, i,j \in \{1,2\},
\end{equation}
with $q_I$ acting as a weight function that causes the integral to converge, $\theta$ are the co-ordinates on the plane of the sky, and $\overrightarrow{\theta}$ is the centre of light. From this the $e^{(s)}$ can be found via, 
\begin{equation}
e^{(s)}=\frac{Q_{11}^{(s)}-Q_{22}^{(s)}+2iQ_{12}^{(s)} }{Q_{11}^{(s)}+Q_{22}^{(s)}},
\label{eqn:chi}
\end{equation}
and therefore using equation \ref{eqn:bayesian} find the chi-squared. This requires full knowledge of the size of the galaxy in order to obtain correct $a$, $b$ and angle parameters. Failure to do so will cause a bias in the parameter estimation hence why we used \textit{option 7}. Incidentally we found no increase in error by using \textit{option 7}.
}
which transforms the a, b and angle of the galaxy into the complex ellipticity $e^{(I)}$ in the image plane using
\begin{equation}
e^{(I)}= \frac{a^2-b^2}{a^2+b^2}\exp(2i\theta),
\label{e_image}
\end{equation}
\noindent removing any information on the size of galaxy.
The ellipticity $e^{(I)}$ is then transformed into the source plane via in the inverse of equation (\ref{eqn:e_image}).
From this the chi-squared calculated in equation \ref{eqn:bayesian} is made.  Since {\tt Lenstool} calculates the source ellipticity in this way we assign some nominal value to the size of the galaxy in the simulations.

Weak lensing mass reconstructions inevitably have limited resolution, because shear is a non-local effect (see equations~\ref{eqn:convergencedefn}--\ref{eqn:sheardefn}) and because the shear field is sampled only at the positions of a finite number of background galaxies.

Fortunately, all that is required to get a robust measurement of bulleticity, is unbiased centroid measurements.
Where available, strong lensing dramatically tightens the resolution of mass maps -- but to rely on strong lensing would unacceptably reduce the number of clusters that we could use.

Since bulleticity measurements will always require overlapping X-ray observations, we will use them to inject information into the reconstruction as a Bayesian prior. 
By assuming that each X-ray peak has an associated group of galaxies, and that the maximum signal for bulleticity is $\sim10\arcsec$ at a redshift of 0.1 (MKN), it is only necessary to consider mass peaks within a small area around the substructure and main cluster only, ignoring the rest of the field of view. 
So if the X-ray suggests a two body configuration then this is the model that is used.

\begin{figure*}
	\begin{minipage}{160mm}
		\begin{centering}
	 		\subfloat[Main halo positional estimates]{\label{fig:single1}\includegraphics[width = 5.5cm]{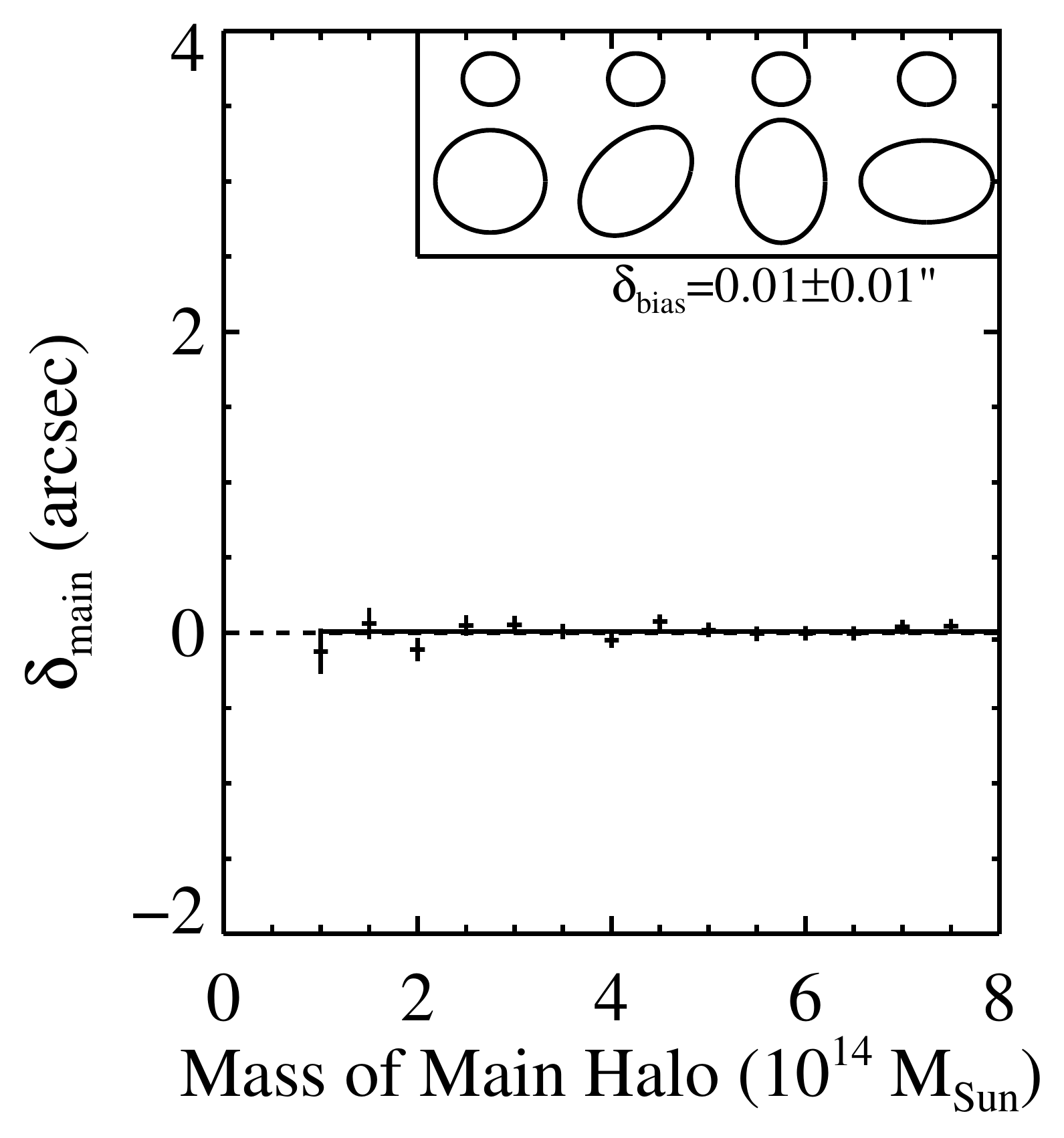}}
			\qquad
			\subfloat[Sub-halo positional estimates]{\label{fig:single2}\includegraphics[width = 5.5cm]{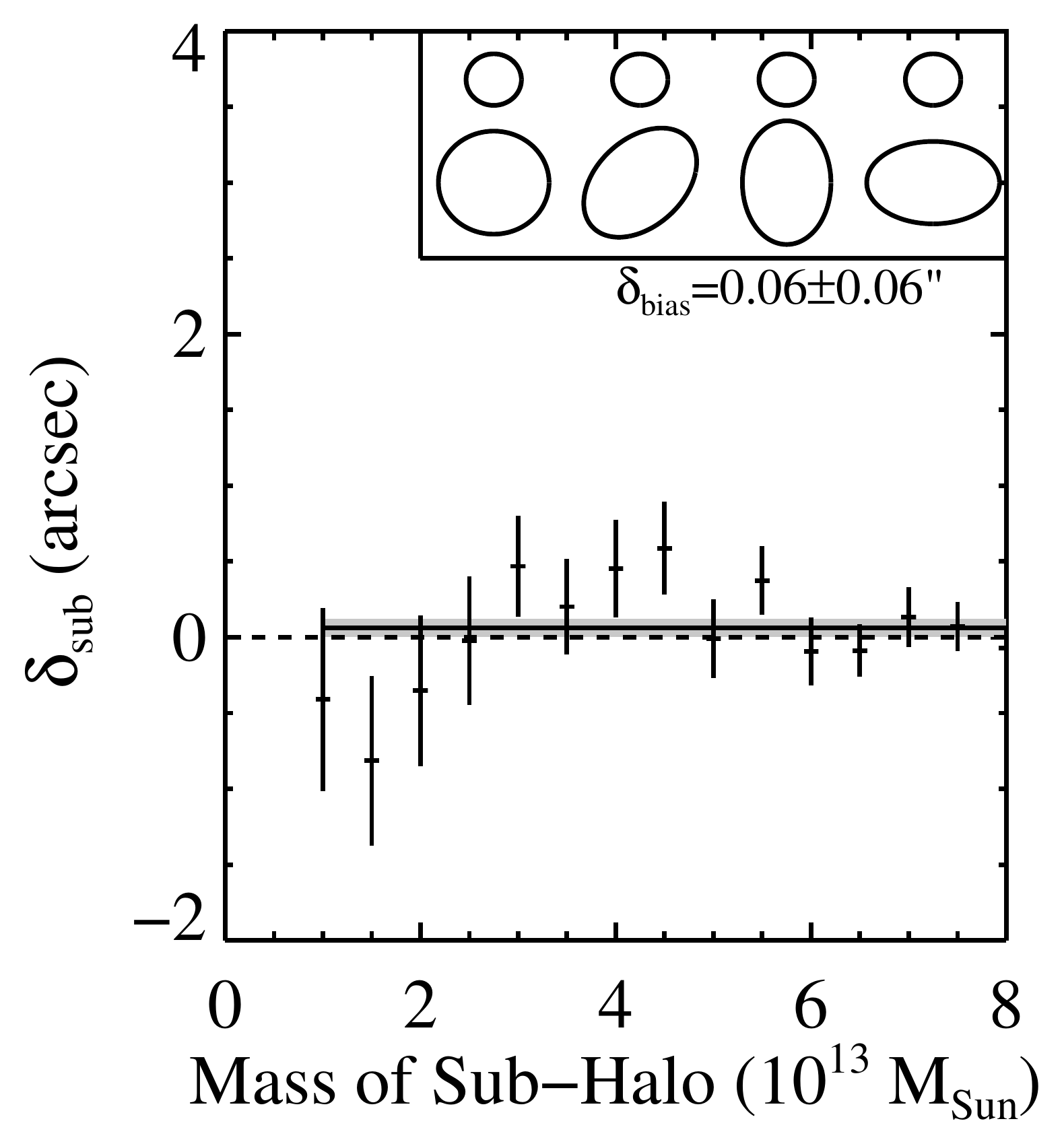}}
			\caption{\label{fig:single}
			\textbf{Intrinsic ellipticities only}: Figure \ref{fig:single1} and \ref{fig:single2} show the positional estimates of the main and sub-halo respectively.  In this initial test the background galaxies only contained a Gaussian intrinsic ellipticity distribution. The mass of the respective halos are shown, in all cases the sub-halo was 10 times less massive than the main halo. (So results at $8\times10^{14}M_\odot$ main halo in figure \ref{fig:single1} are from the same simulation as those shown at $8\times10^{13}M_\odot$ in figure \ref{fig:single2}). A variety of configurations were tested with the cartoon inset showing the setup in each case. For each configuration, 100 noise realisations were run at the mass scale, the position of each halo estimated and then averaged over the all configurations. (so each point reflects 400 averaged simulations). }
			\end{centering}
		\end{minipage}	
\end{figure*}

Table~\ref{tab:priors} shows the positions and masses of the clumps simulated with the associated priors used.

\subsubsection{Estimation of Sub-Halo Position}\label{sec:bias}

In order to understand {\tt Lenstool} and its behaviour in the weak lensing limit for a two halo system, we tested it on noise free simulations where the galaxies were inherently circular and the only affect was gravitational shear. Since Lenstool is a maximum likelihood algorithm in the case of zero noise the chi-squared calculation becomes undefined, so therefore we set the variance of ellipticity in {\tt Lenstool} to a very small value ($0.01$).

Figure \ref{fig:asymm} shows the full posteriors for the positions of the main and sub-halo.
The positions from the sampler have been binned with the maximum likelihoods shown as solid lines and the true values as dotted.
The top panels show radial and tangential position of the main halo and the lower panels show the radial and tangential positions of the sub-halo. 
It is clear that in the situation where there is no noise and the exact profile is given to lenstool the maximum likelihood is centred on the true position with extremely small variance.

\begin{figure*}
	\begin{minipage}{180mm}
		\begin{centering}
			\subfloat[Main halo positional estimates]{\label{fig:shape1}\includegraphics[width = 5.5cm]{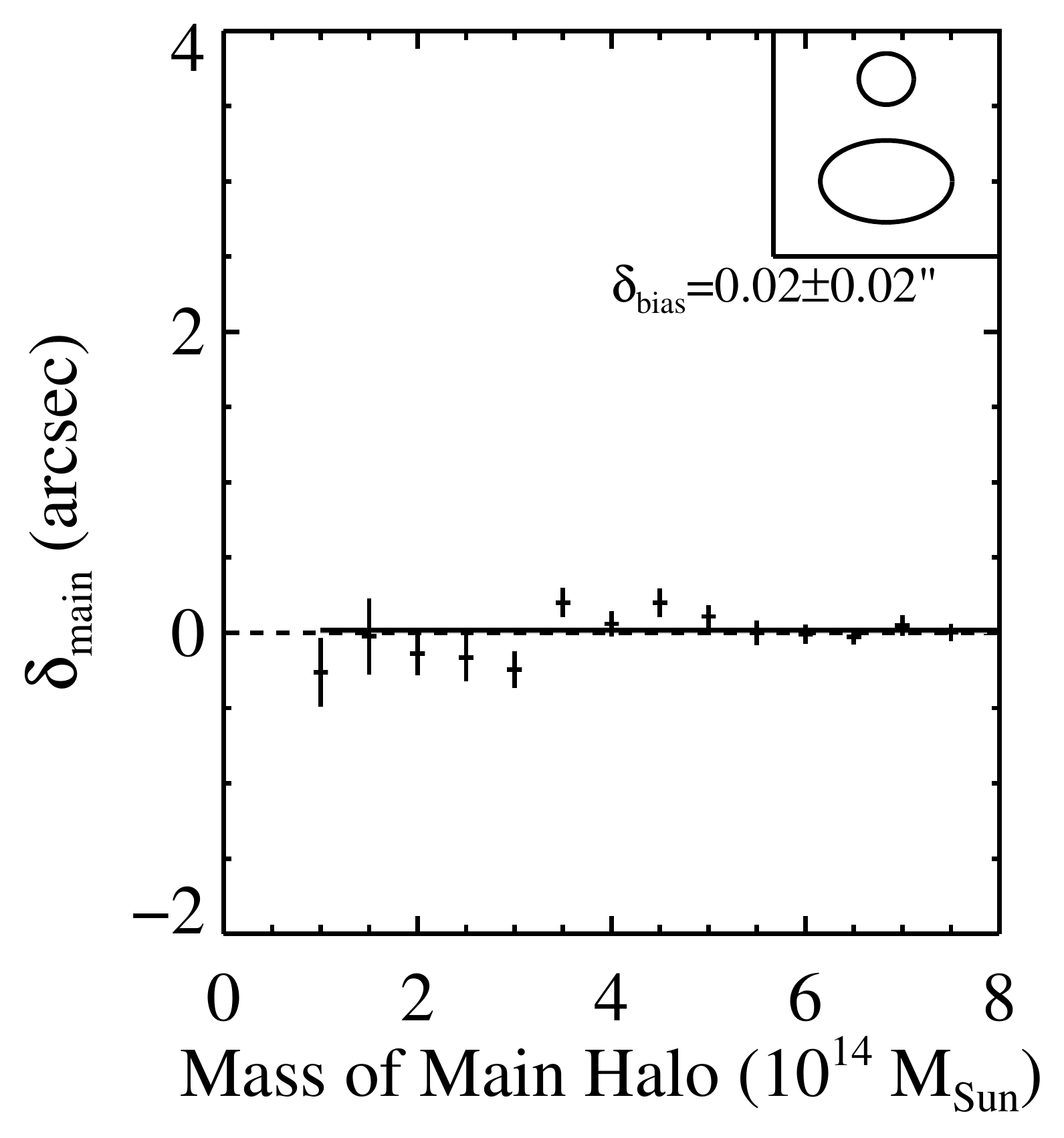}} 
			\subfloat[Sub-halo positional estimates]{\label{fig:shape2}\includegraphics[width = 5.5cm]{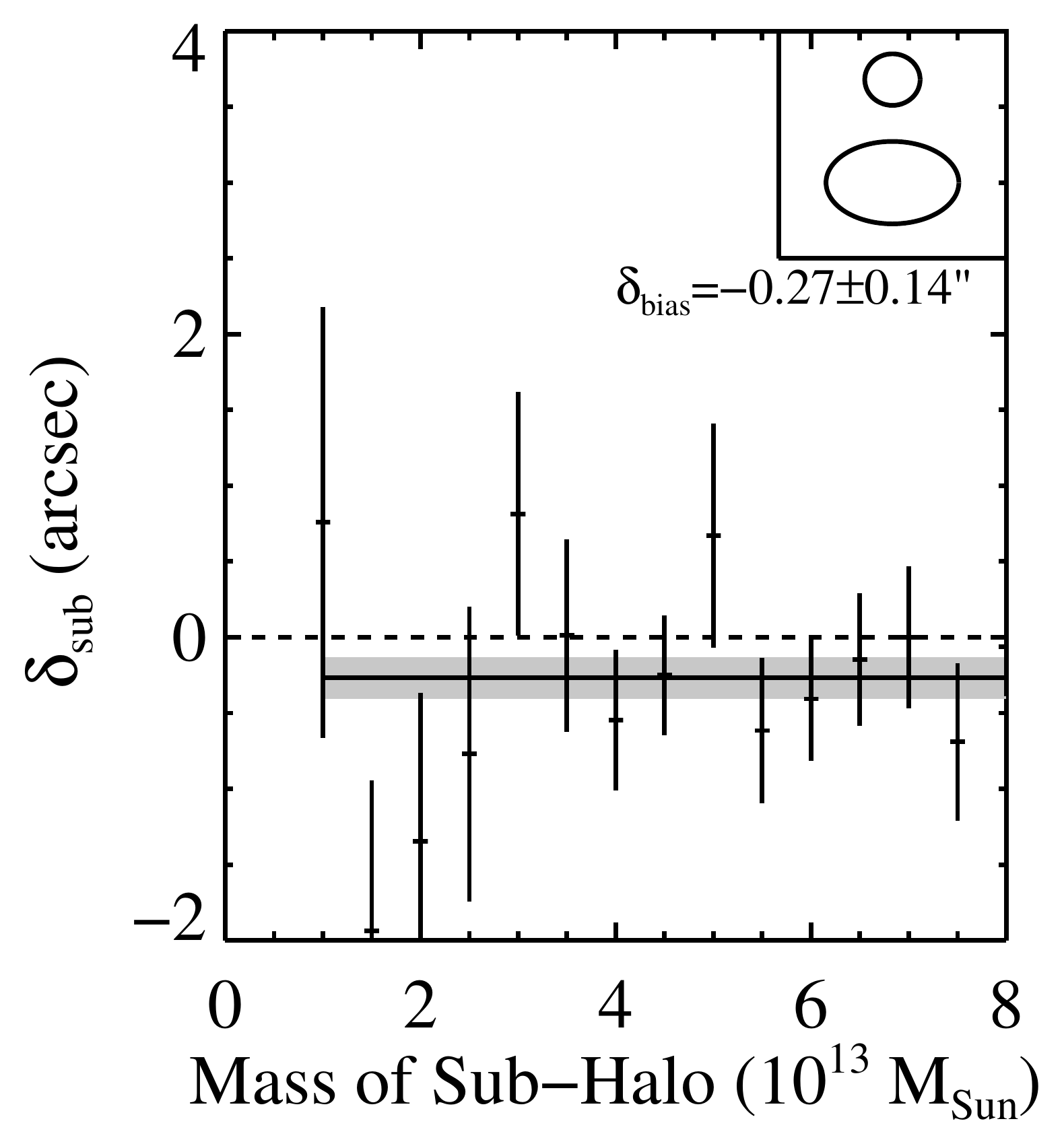}} 
			\caption{\label{fig:shape}
			\textbf{Intrinsic ellipticities and  shape measurement bias}: Figure \ref{fig:shape1} and \ref{fig:shape2} are the positional estimates of the main and sub-halo respectively. In each case the mass is given and the main halo is 10 times more massive than the sub-halo. The main halo is elliptical and the background galaxies have shape measurement bias and intrinsic ellipticities. In this case 100 realisations were run and the average position at each mass scale calculated.}
		\end{centering}
	\end{minipage}
\end{figure*}
\begin{figure*}
	\begin{minipage}{180mm}
		\begin{centering}
			\subfloat[Main halo positional estimates]{\label{fig:rshift1}\includegraphics[width = 5.5cm]{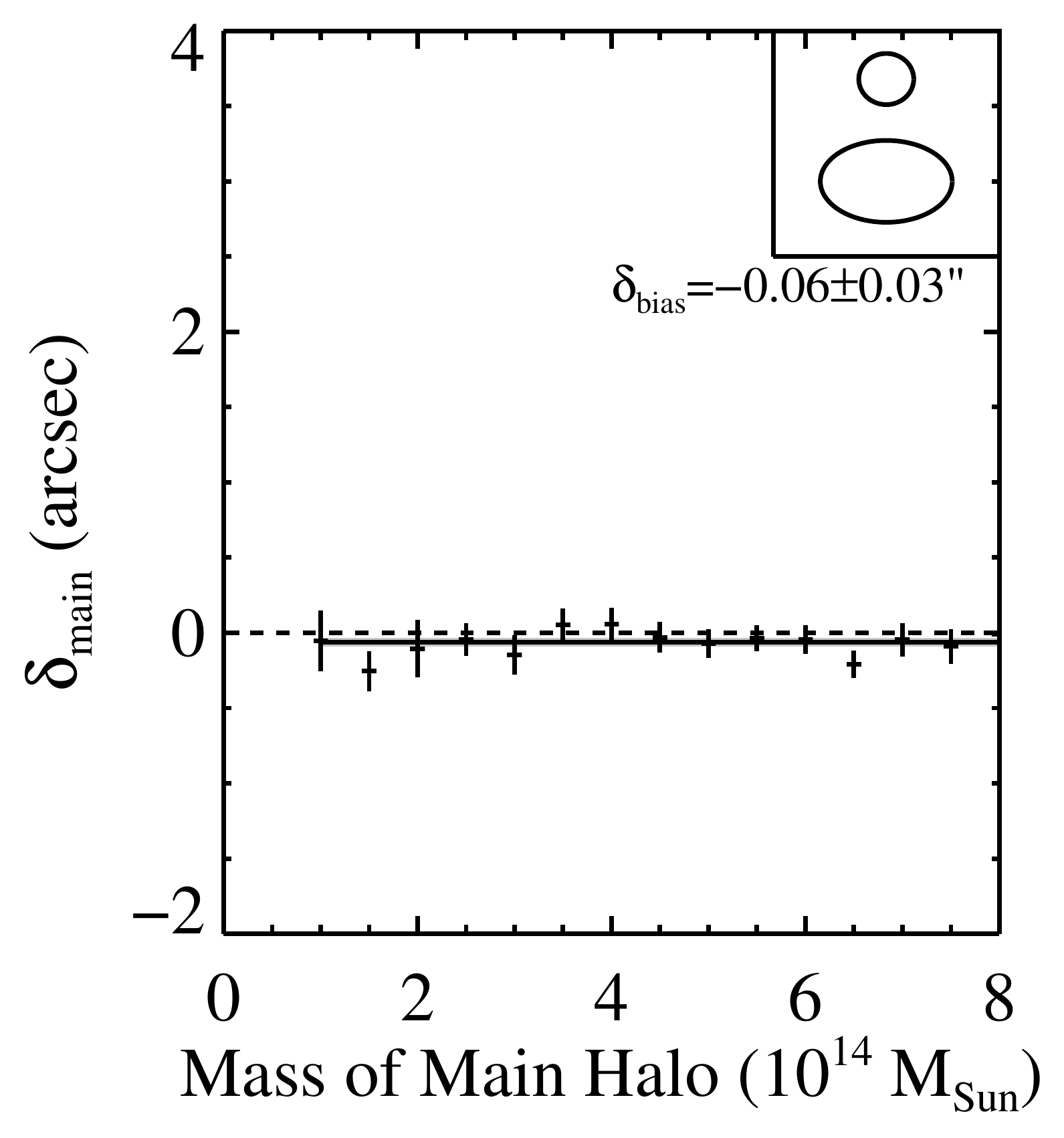}} 
			\subfloat[Sub-halo positional estimates]{\label{fig:rshift2}\includegraphics[width = 5.5cm]{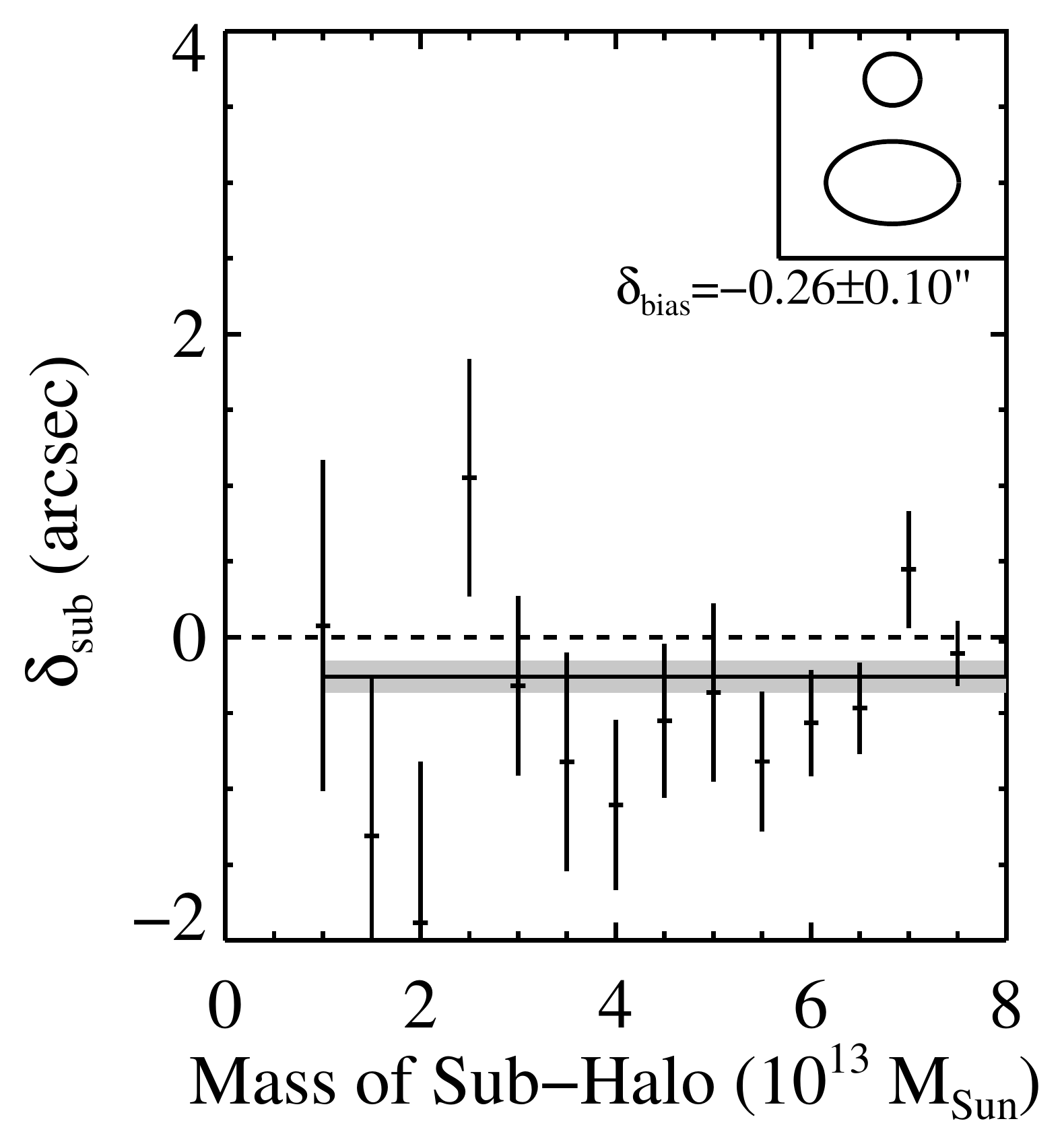}} 
			\caption{\label{fig:rshift}
			\textbf{Intrinsic ellipticities, shape measurement bias and source galaxy redshift distribution}: Figure \ref{fig:shape1} and \ref{fig:shape2} are the positional estimates of the main and sub-halo respectively. In each case the mass is given and the main halo is 10 times more massive than the sub-halo. The main halo is elliptical and the background galaxies have shape measurement bias, a distribution in their redshift and intrinsic ellipticities. In this case 100 realisations were run and the average position at each mass scale calculated.}
		\end{centering}
	\end{minipage}
\end{figure*}

\section{Results}\label{sec:results}

The expected offset between dark and baryonic components is $\sim2\arcsec$ ($\sim3.5\arcsec$)  at a redshift of 0.6 (0.3) at a radial of distance of $0.15r_{500}$ and therefore any bias needs to be subdominant in comparison. 
The redshift distribution of clusters in the COSMOS field (\citealt{2007ApJS..172..182F}) suggest that we expect a similar number of clusters at a redshift of $0.3$ to $0.6$, therefore the measurement of an offset can tolerate $\sim0.5\arcsec$ bias in the reconstruction in order to measure a bulleticity signal to $\sim3\sigma$ significance detection.

In an attempt to understand the behaviour of {\tt Lenstool}, initial simple simulations were run and then an increasing number of contaminants and complexities were introduced. 
Unless stated otherwise, each panel in each figure shows $\delta$: the maximum likelihood radial position minus the true position for a given mass and simulation configuration, weighted averaged over 100 realisations for a given mass scale and then averaged over each configuration shown in the cartoon inset of the plot, i.e.

\begin{equation}
\delta_{\rm halo}(m)=\langle\langle r_{\rm Meas}(m) - r_{\rm True} \rangle_{100}\rangle_{\rm config},
\end{equation}

\noindent and the error for the given configuration is just given by the error in the mean. Furthermore the main halo is always 10 times more massive than the sub-halo (and appropriate concentrations given by equation \ref{eqn:conc}).

\begin{figure*}
	\begin{minipage}{160mm}
		\begin{centering}
  			\subfloat[Main Halo]{\label{fig:multiple1}\includegraphics[width = 5.5cm]{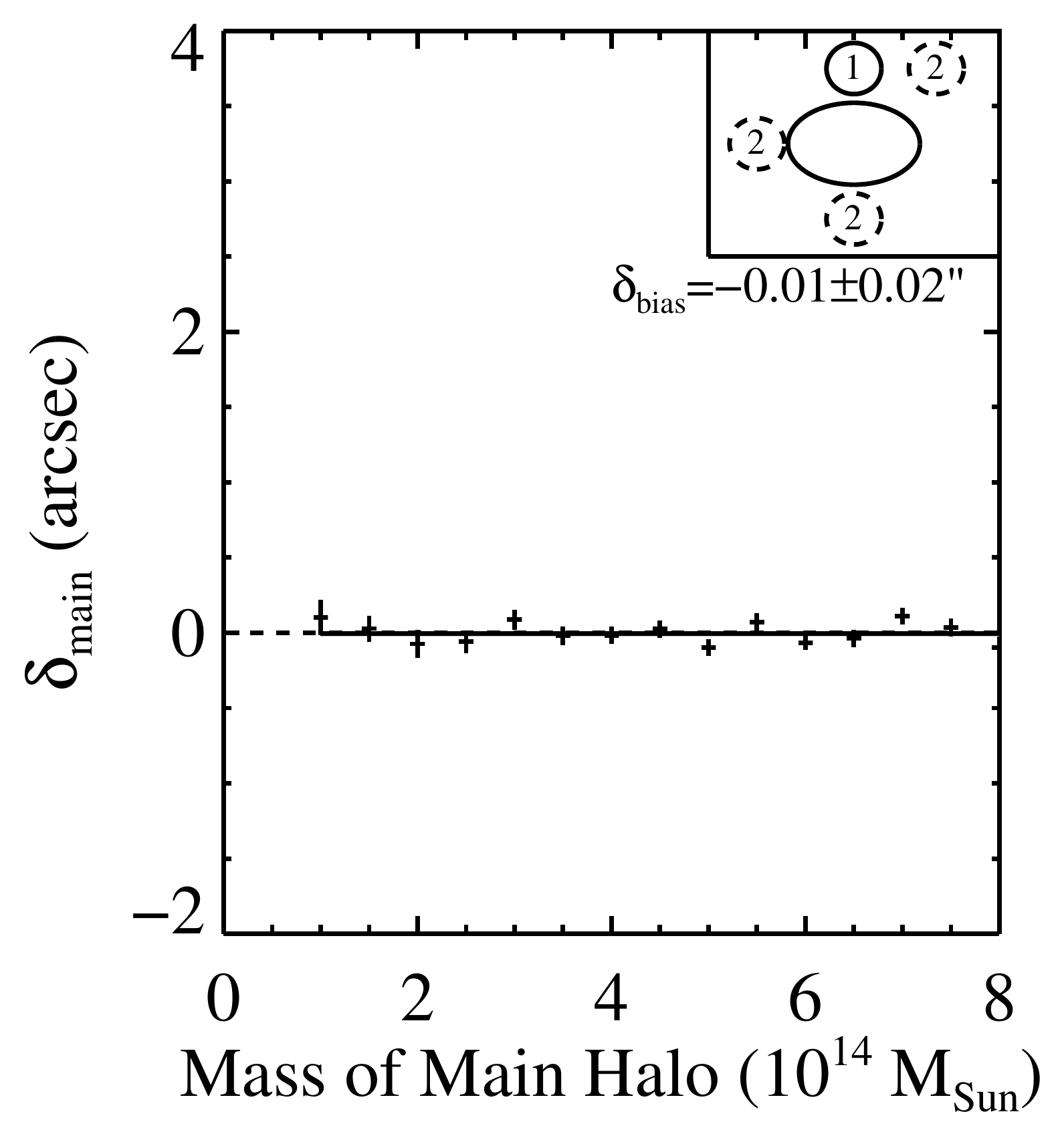}} 
			\subfloat[Sub-Halo 1] {\label{fig:multiple21}\includegraphics[width = 5.5cm]{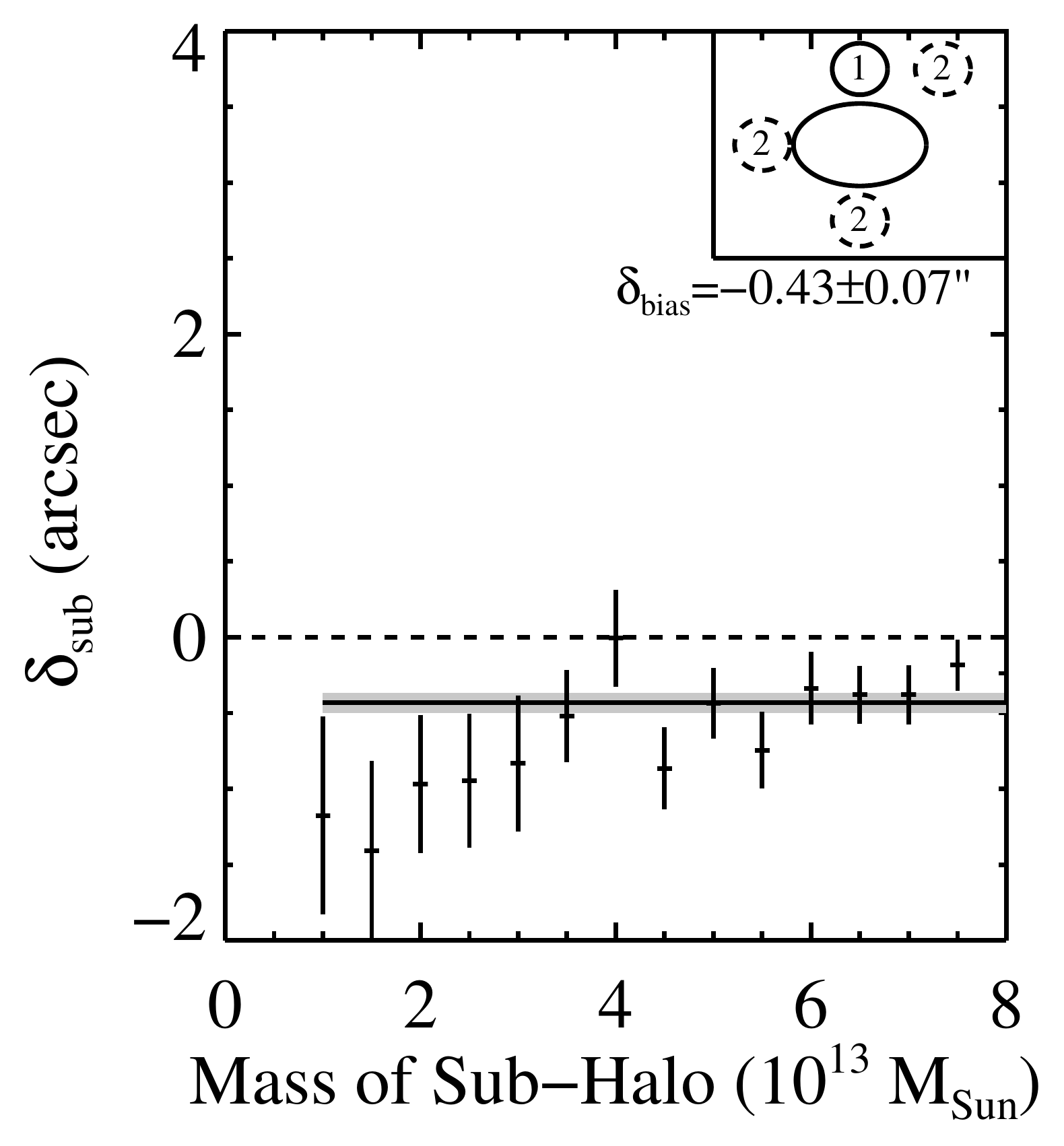}} 
			\subfloat[Sub-Halo2] {\label{fig:multiple22}\includegraphics[width = 5.5cm]{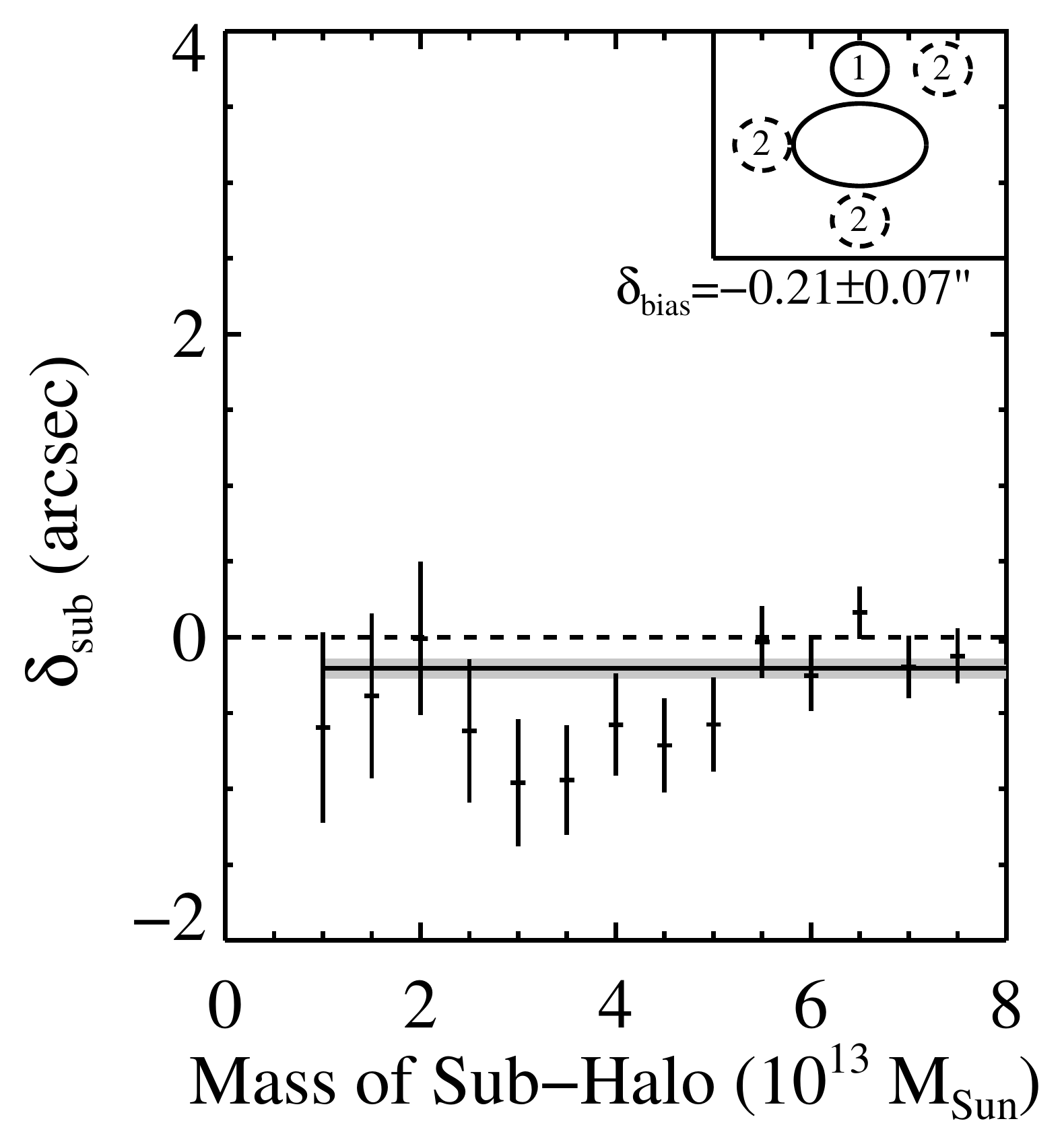}} 
			\caption{\label{fig:multiple}
			\textbf{Dual sub-halo simulation, with intrinsic ellipticities, shear measurement bias and source galaxy redshift distribution}: Figure \ref{fig:multiple1} shows the positional estimates of the main halo, figure \ref{fig:multiple21} the estimates of \textit{sub-halo 1} and \ref{fig:multiple22} gives the estimates of \textit{sub-halo 2}. The masses of the halos are given. In each case the sub-halos are 10 times smaller in mass (so they are equal size) than the main halo. The background galaxies have intrinsic ellipticities, shape measurement bias and a redshift distribution. The plots show 3 different configurations (given by the dashed circles). In each case \textit{sub-halo 1} is kept in the same place as shown in the cartoon inset, and for each of the 3 scenarios \textit{sub-halo 2} is positioned as shown. In each scenario 100 noise realisations are run and the positions averaged over all configurations and noise realisations.}
		\end{centering}
	\end{minipage}
\end{figure*}

\begin{figure*}
	\begin{minipage}{160mm}
		\begin{centering}
    			\subfloat[Main  halo positional estimates]{\label{fig:SIS1}\includegraphics[width = 5.5cm]{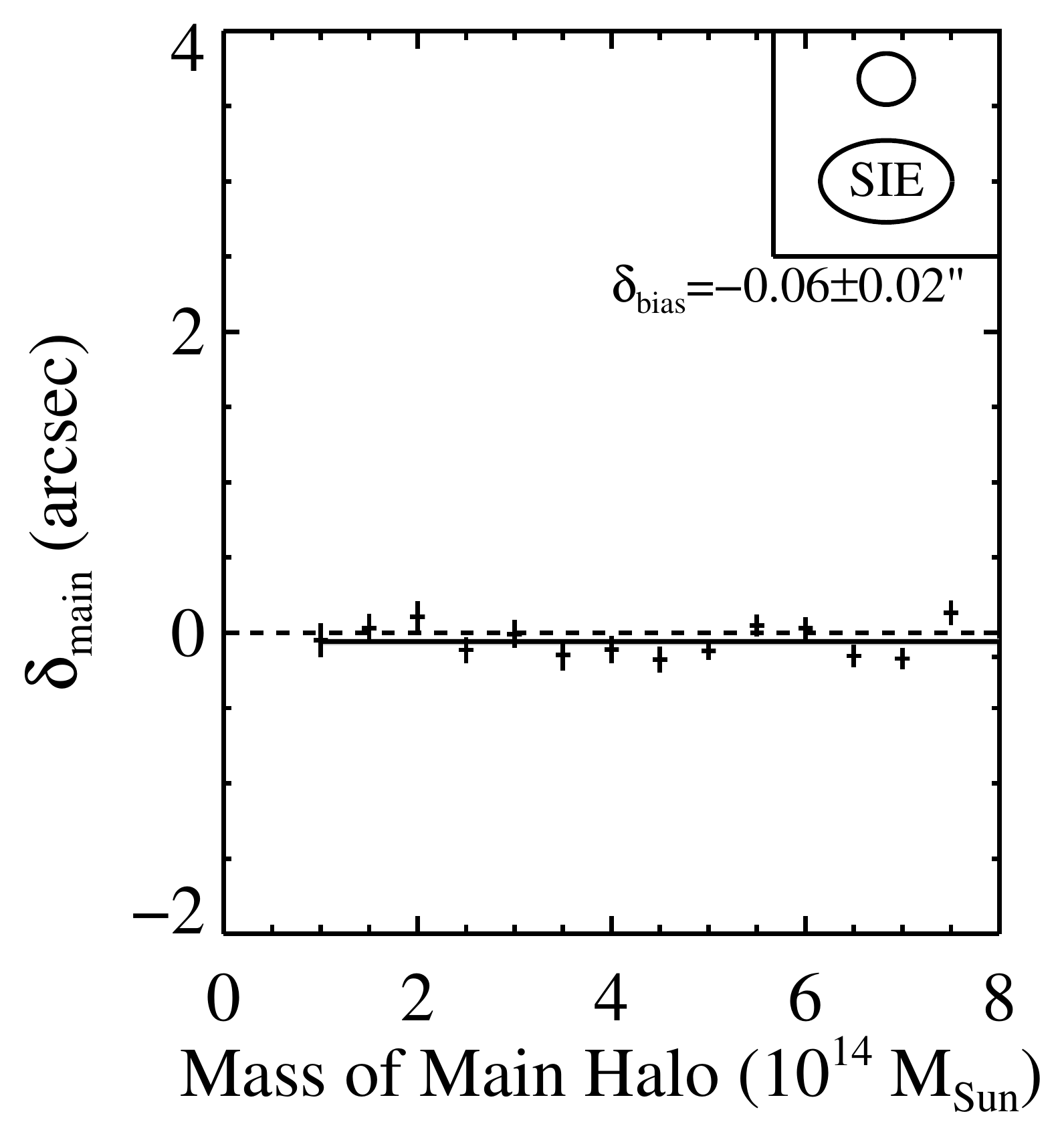}} 
			\subfloat[Sub-halo positional estimates]{\label{fig:SIS2}\includegraphics[width = 5.5cm]{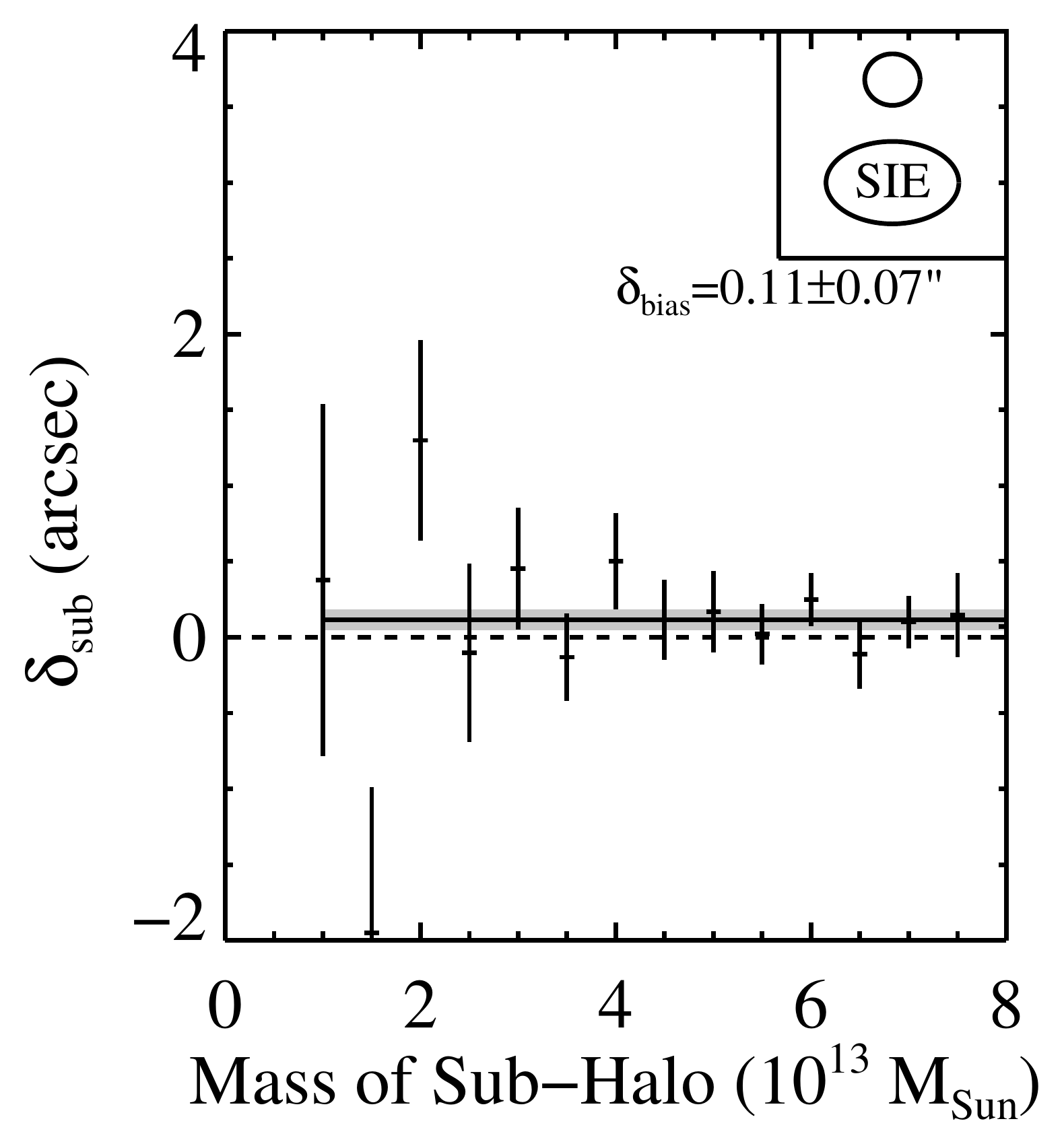}} 
			\caption{\label{fig:SIS}
			\textbf{ Simulated SIE halos, with intrinsic ellipticities, source galaxy redshift distribution and shape measurement bias}: Figures \ref{fig:SIS1} and \ref{fig:SIS2} show  the main and sub-halo positional estimates respectively. In this scenario a SIE profile is simulated and NFW fitted imitating profile misidentification in real data. The main halo is always 10 times larger than the sub-halo, and the source galaxies have shape measurement bias, intrinsic ellipticities and a redshift distribution. 100 noise realisations were run and the average position estimated in each case. }
		\end{centering}
	\end{minipage}
\end{figure*}

We carried out four initial tests in the simplest two body case: one with a circular main halo and three with elliptical main halos at different angles. Each test contained a simple background galaxy Gaussian intrinsic distribution and was run 100 times with different noise realisations. We then fitted two lines of best fit to the data to determine any significant bias in the positional estimates. One was a constant offset and the other a mass dependant one. 
For the sub halo we found that the reduced chi-square for a mass dependant line was 1.34, whereas for a constant offset we found a chi-square of 1.25. 
Thus we found no significant evidence for a mass dependant bias and therefore fitted a constant offset.

Each configuration showed in the cartoon inset of Figure \ref{fig:single} exhibited no bias and therefore in order to better constrain the error on positional estimates we compiled the results into Figure \ref{fig:single} giving the combined results from the initial tests.

It was found that {\tt Lenstool} was robust to a basic level of noise so we introduced further sources of contaminants. Figures \ref{fig:shape} shows $\delta$ for the main and sub-halo respectively when shape measurement bias is introduced.
Figure \ref{fig:shape1} seems to show that the the shape measurement bias has no affect on  the positional estimate of the main halo, however the sub-halo in Figure \ref{fig:shape2}, seems to be slightly biased in the negative radial direction (towards the main halo).
The cause of this will be the preferred direction of each galaxy. 
The level of the bias is of order 0.01, which is a similar level to the expected signal from a dark matter sub-halo.
Because each galaxy has a preferred direction it will mean that the preferred fit of the halo will not be the correct one, causing a bias in the position. This bias of $0.27\pm0.14\arcsec$ is well within the tolerated level.

Gravitational lensing is a geometrical affect and hence the signal is dependant on the distance the galaxy is from the halo. {\tt Lenstool} requires knowledge of the source galaxy redshift, we therefore introduce a redshift distribution into to source galaxies and test the approximation that their redshifts are 1. Figure \ref{fig:rshift} shows the results when such a distribution is introduced. Figure \ref{fig:rshift2} has no significant evidence for an increase in bias due a source galaxy redshift approximation from that of Figure \ref{fig:shape2}

Using the same signal contaminants as Figure \ref{fig:rshift} (intrinsic ellipticities, source redshift and shape measurement bias), we introduce a second sub-halo, complicating the geometrical setup of the simulations. We ran three different scenarios; in each case the main halo was at the centre of the field and sub-halo 1 was kept at the position previously simulated. The new, second sub halo was placed at three different locations as shown in the cartoon inset. For each sub halo 2 position 100 noise realisations were run. In all cases the main halo was 10 times more massive than the sub-halos, and the sub-halos were equal size. We found that there was no preferred bias dependant on the position of the second sub-halo and so averaged these simulations together in order to better constrain the bias and error bars. Figure \ref{fig:multiple} shows hat the bias introduced by the source galaxy distribution is evident in the two body system. The more complicated geometrical setup seems to have no affect on the overall bias.

Figure \ref{fig:SIS} is a test into the model dependancy of the reconstruction method. 
Although NFW profiles have been extensively studied with both simulated and empirical data, the inclusion of baryons have shown that profiles depart from the assumed NFW (\citealt{2010MNRAS.405.2161D}). 
Therefore in order to understand what the affect of this is, Figures \ref{fig:SIS} show the positional accuracy of the main and sub-halo in the case where a singular isothermal ellipsoid is simulated and using {\tt Lenstool} a NFW is fitted. Figure \ref{fig:SIS2} shows that the sub halo has an insignificant positive bias, seemingly in contradiction to previous results. 

This unbiased nature is due to the fact that the central core of an SIS is extremely peaked. An NFW has a much flatter profile in the core and therefore may introduce more uncertainty in the peak position. 
One affect of introducing an SIS is that the scatter seems to be much larger and at smaller masses the positional estimates become unreliable.

\begin{figure*}
		\begin{minipage}{180mm}
		\begin{centering}
	 		\subfloat[Main Halo]{\label{fig:functr_1}\includegraphics[width = 4.5cm]{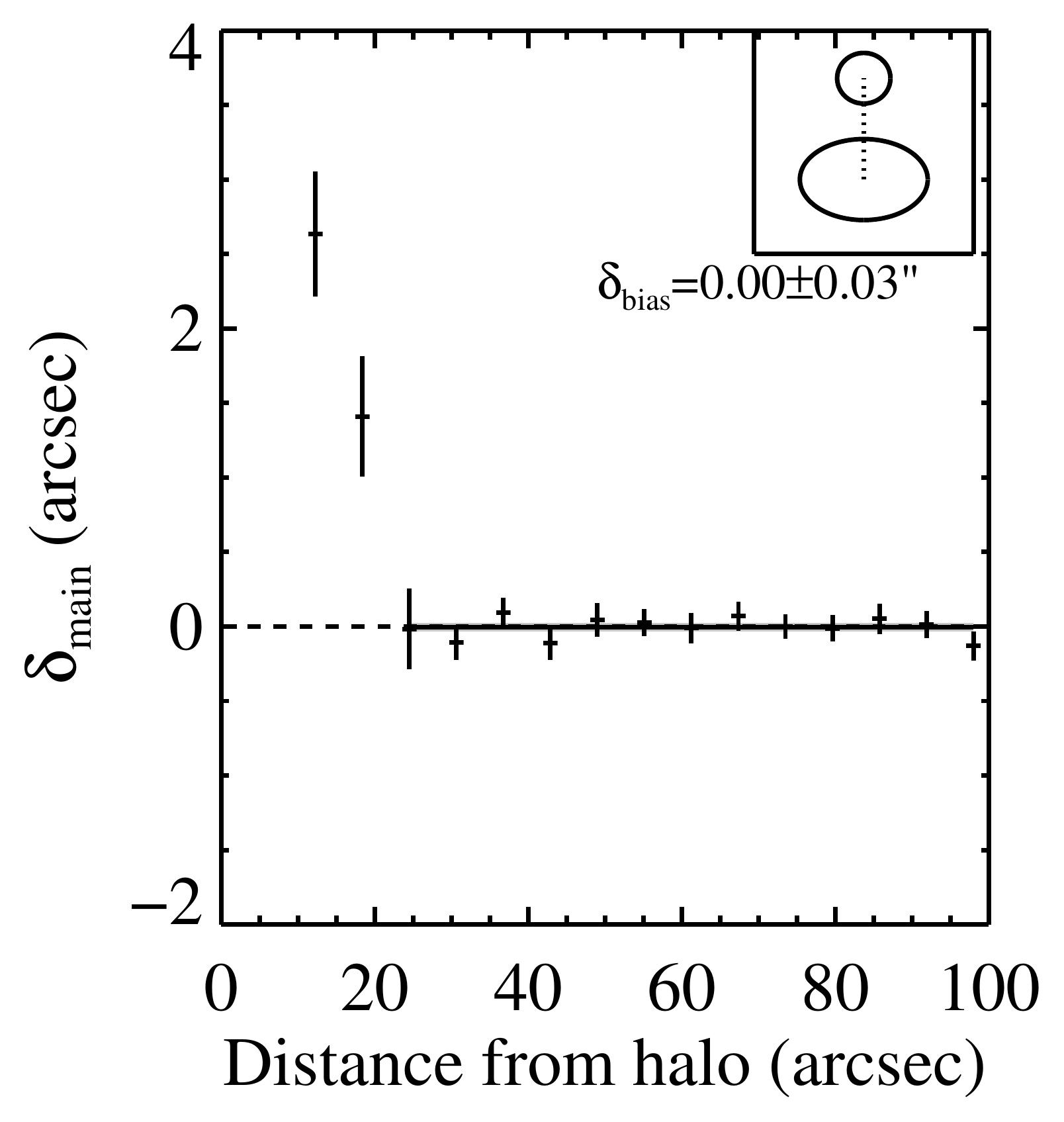}}
			\subfloat[Sub Halo]{\label{fig:functr_2}\includegraphics[width = 4.5cm]{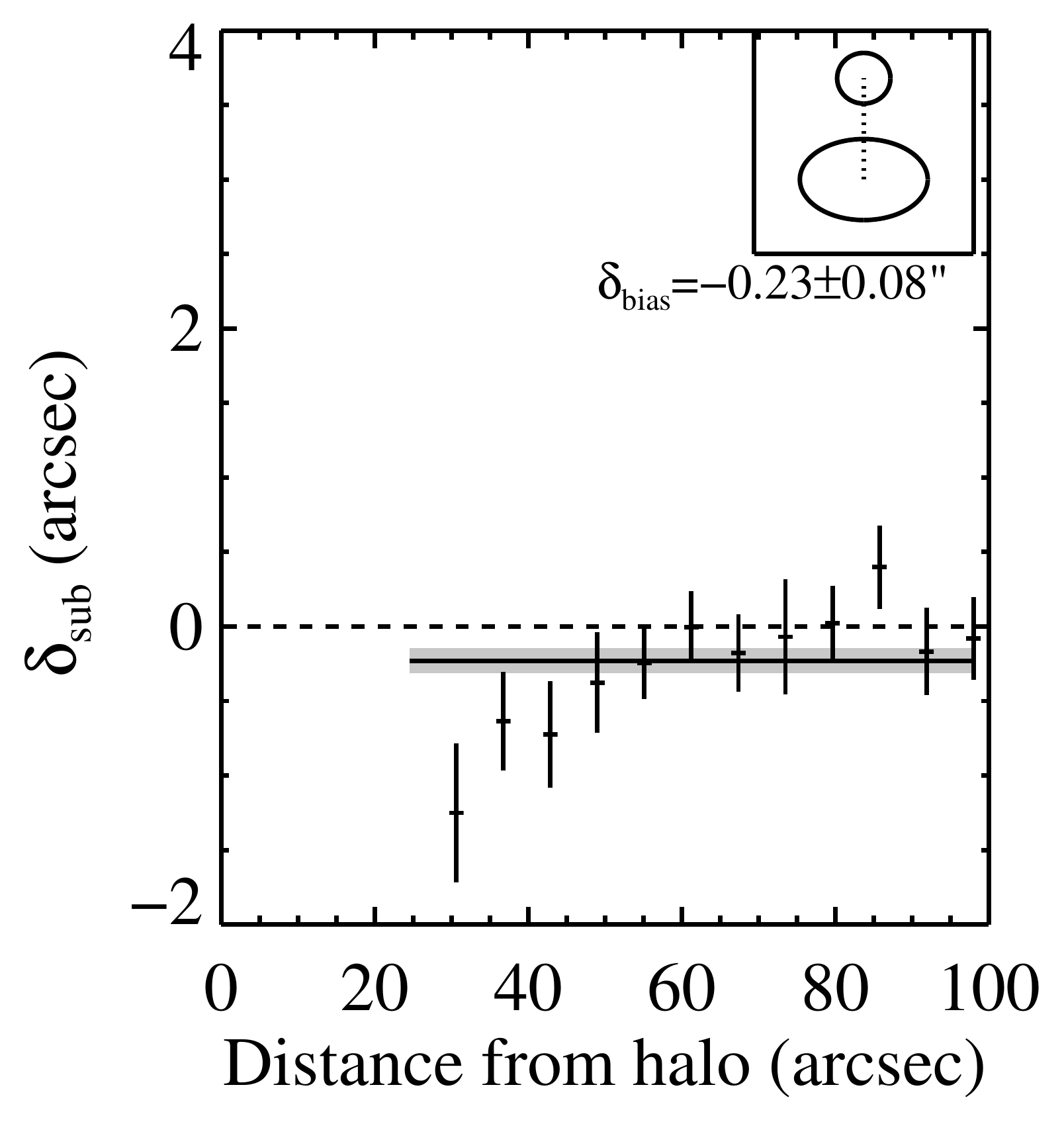}}
			\subfloat[Main Halo]{\label{fig:frac_1}\includegraphics[width = 4.5cm]{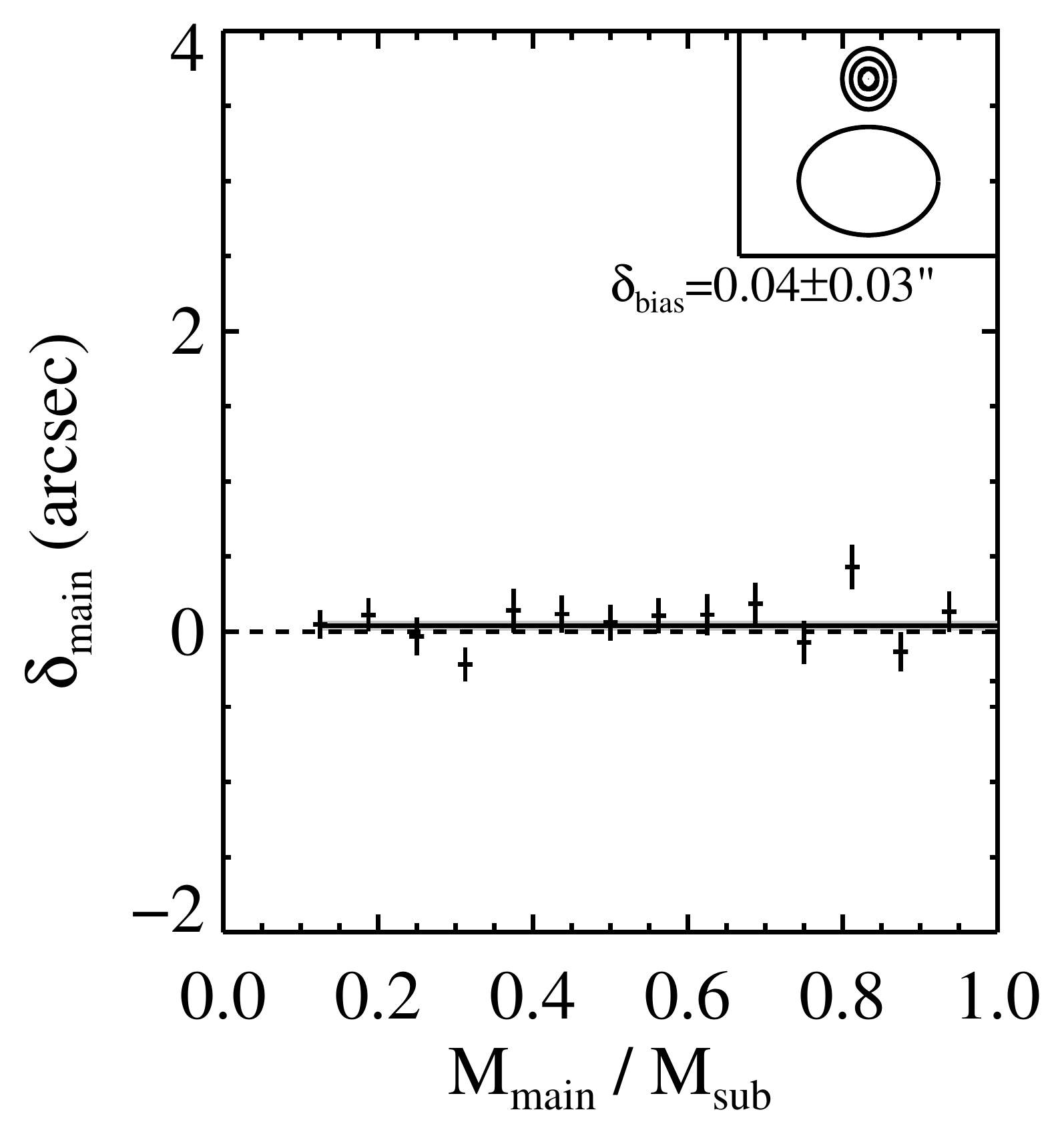}}
			\subfloat[Sub Halo]{\label{fig:frac_2}\includegraphics[width = 4.5cm]{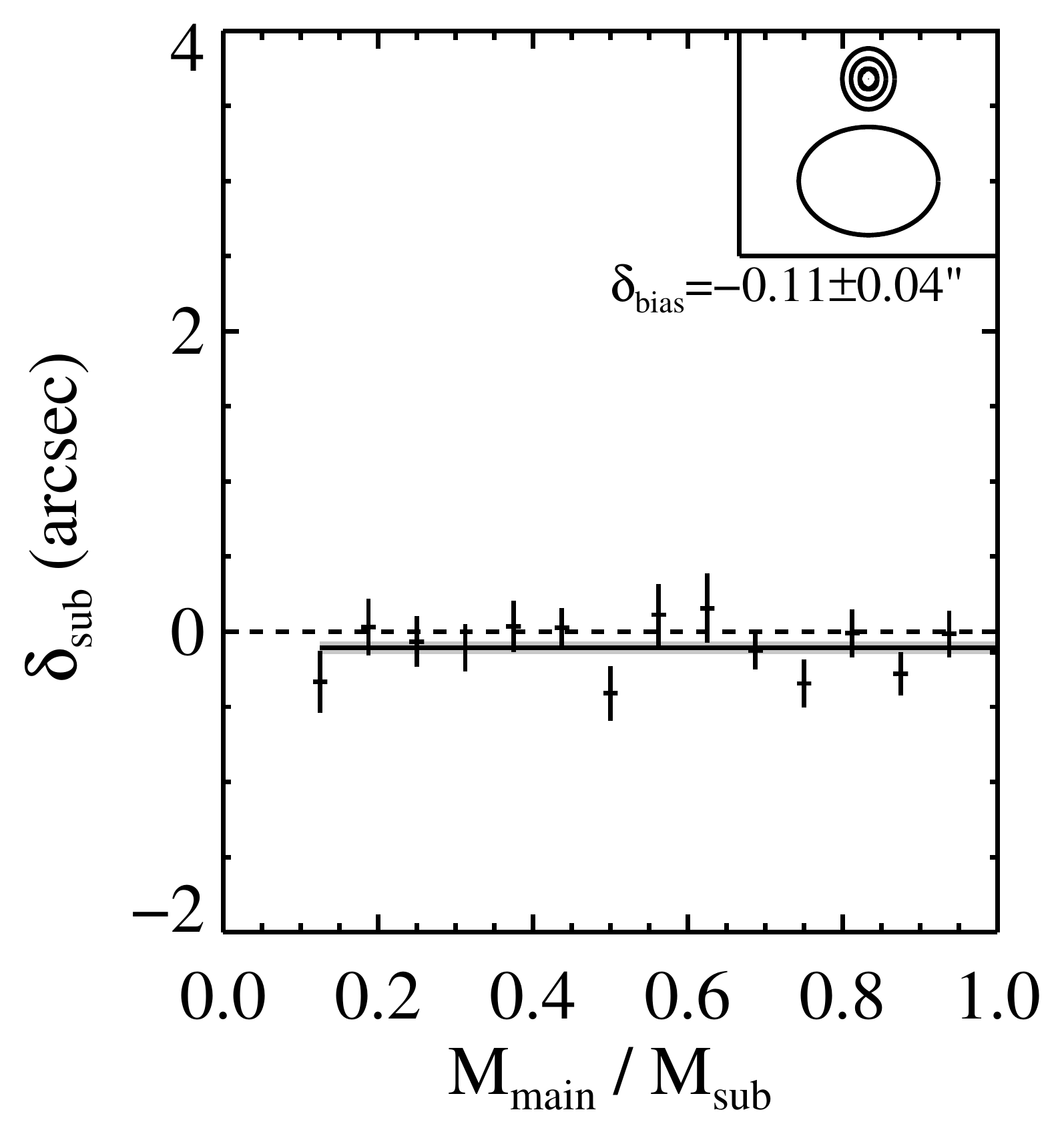}}
		\caption{\label{fig:functr}
			\textbf{Weak lensing accuracy as a function of the radial position from the cluster and mass fraction ($\sfrac{M_{\rm sub}}{M_{\rm main}}$)}. In each case the background galaxies have intrinsic ellipticities, redshift distribution and shape measurement bias. The first two panels (main and sub-halo respectively) show an $8\times10^{13}M_\odot$ cluster (with associated 10 times large parent halo), simulated at various distances from the main halo. The catastrophic failure at $<30\arcsec$ is due to the sub-halo position overlapping with the parent halo. The second two panels show  the positional estimates of a main halo of $8\times10^{14}$ and sub-halo with an increasing sub-halo mass (decreasing ratio). It is shown that the bias is mass fraction independent and is robust to minor mergers as well as substructure infall.}
		\end{centering}
	\end{minipage}
\end{figure*}


\begin{figure*}
	\begin{centering}
		\subfloat{\label{fig:cont1}\includegraphics[width = 5.5cm]{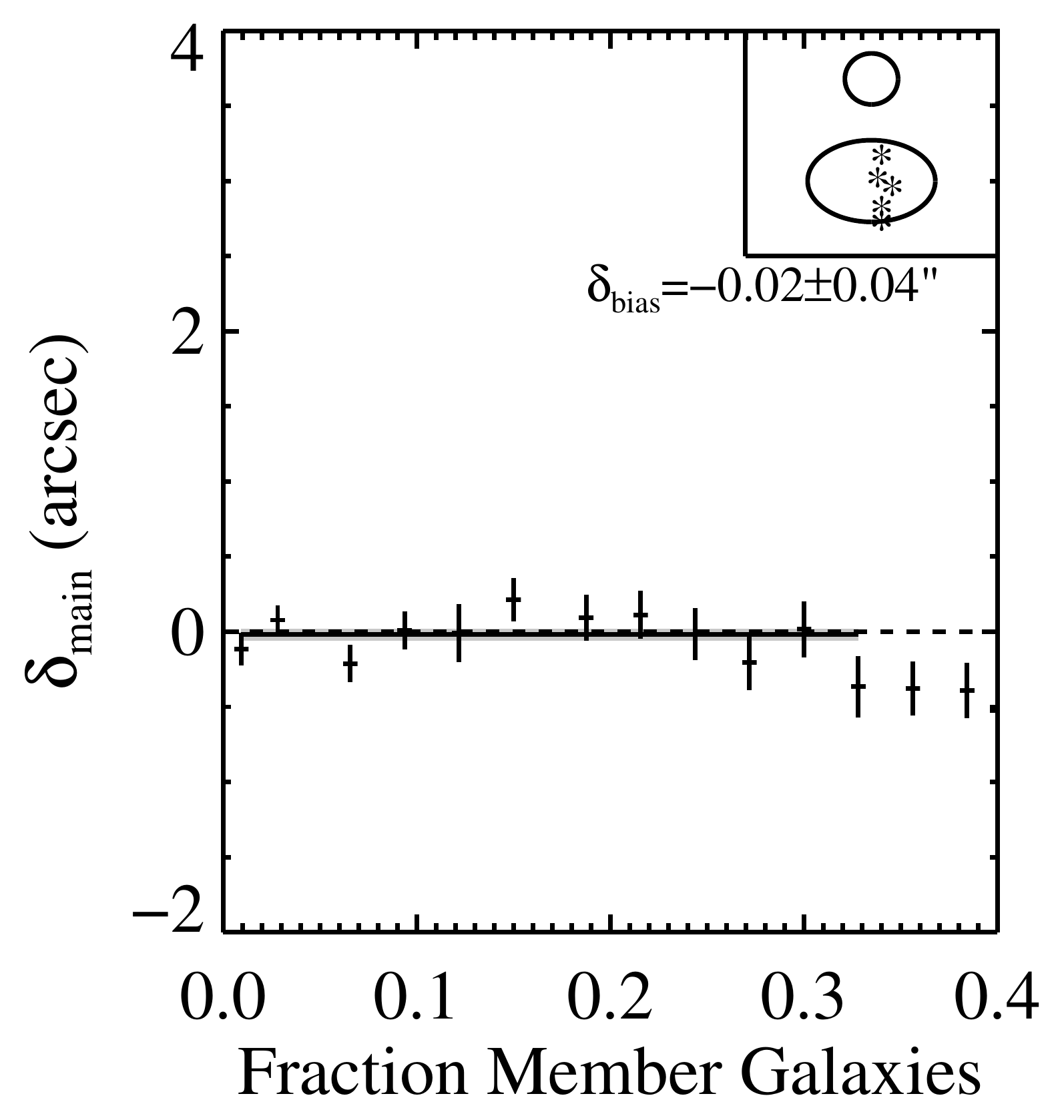}}
  		\subfloat{\label{fig:cont2}\includegraphics[width = 5.5cm]{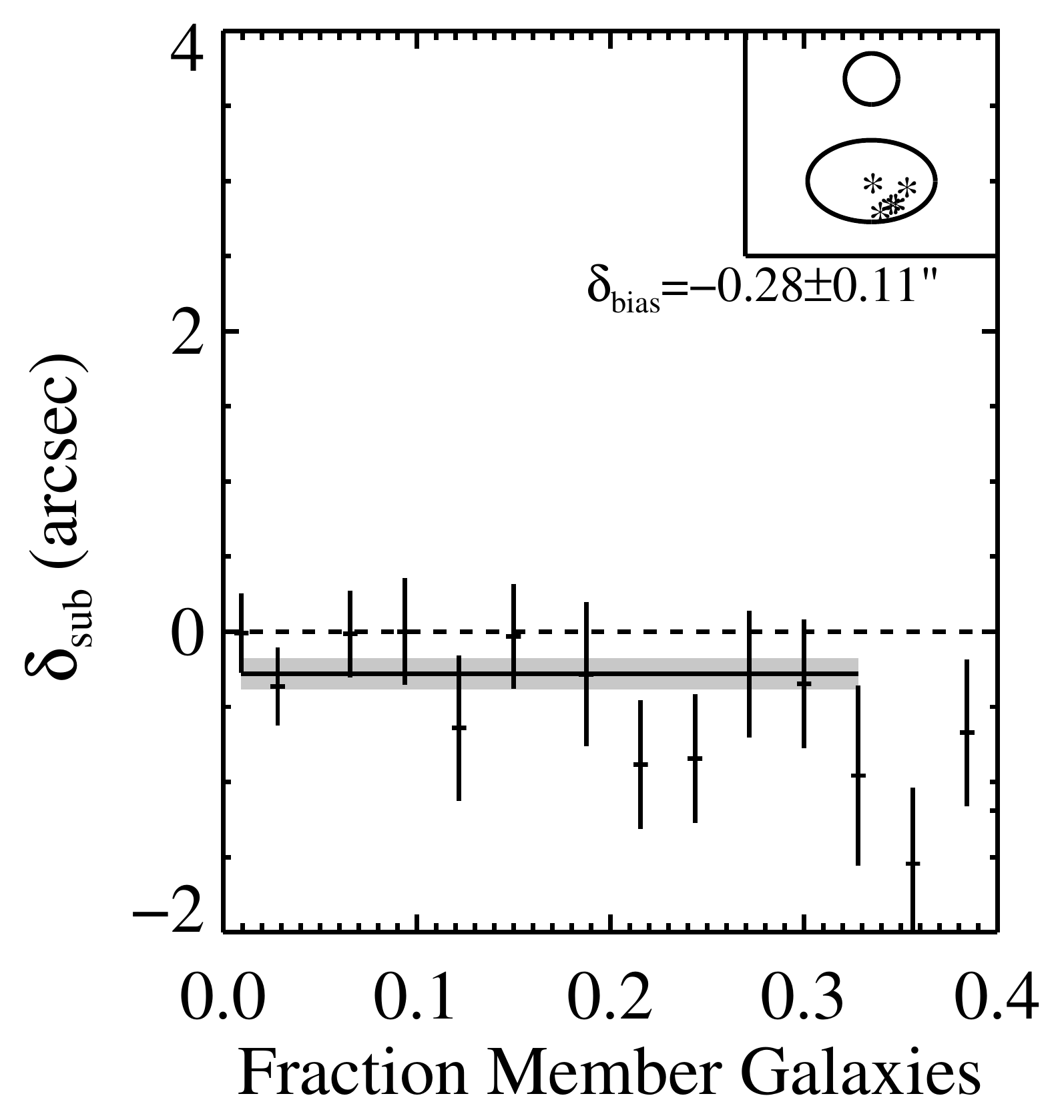}} 
 		\caption{\label{fig:cont}
 		Main (top) and Sub (bottom panel) position as function of cluster member contamination expressed as a percentage of the total background galaxy number.}
	\end{centering}
\end{figure*}

\subsection{Accuracy as a function of distance from the cluster and mass fraction}
Figure \ref{fig:functr} shows the results if the halo masses are kept constant ($8\times10^{14} M_\odot$) at a redshift of 0.6 with an ellipticity of $0.2$, positional angle of $180^\circ$, shape measurement bias and a background galaxy redshift distribution, and (first two panels) the sub-halo is moved from close to the cluster outwards and (second two panels) the mass fraction is increased ($\sfrac{M_{\rm sub}}{M_{\rm main}}$).

The fitted lines shows over what mass interval the mass independent-bias remains, until the chi-square of the line becomes greater than one standard deviation from the expected value.
Figure \ref{fig:functr} shows that the fit breaks down at low radii ($<30\arcsec$). 
This value coincides with the size of the prior around the sub-halo and shows that the sampler cannot de-merge the two halos.
Figure \ref{fig:frac_1} and \ref{fig:frac_2} show that the bias is independent of mass ratio, and even in the case where the two halos are of equivalent size the bias remains negligible.
This is promising as it shows the reconstruction should be reliable even in the case of a minor merger and not just substructure infall.

\subsection{Cluster Member Inclusion}

Figure \ref{fig:cont} show the results from including member galaxies into the reconstruction. The fraction of background galaxies is calculated by summing the total number of member galaxies in the field of view and dividing by the number of background galaxies. 
It can be seen that the reconstruction is reliable up to $\sim30\%$ of the background galaxies, at which the position of the sub and main halo become unreliable. Given a background density of 80 per sq. arc minute, which is significantly less found by \citet{2011ApJ...726...48H}, we can conclude from these plots that we not worried about inclusion of these member galaxies.

\section{Precision of Lenstool}

Throughout this investigation we have consistently found that by averaging many clusters together one can measure the position of DM halos accurately to within $\sim0.3\arcsec$. 
Understanding how precisely we can measure these offsets informs us how many offsets will need to be measured in order to make a statistically significant detection. 

The error bars derived from {\tt Lenstool} give a rough estimate of the number of sub-halos that are required to robustly measure a significant offset between dark matter and gas. We decide to use the error bars from Figure \ref{fig:rshift} to derive the precision. These were the error bars in the case of all signal contaminants. We found no evidence for additional uncertainty due to member galaxies at the expected level, and therefore have not factored these into the errors.

\begin{figure}
 \includegraphics[width = 8.5cm]{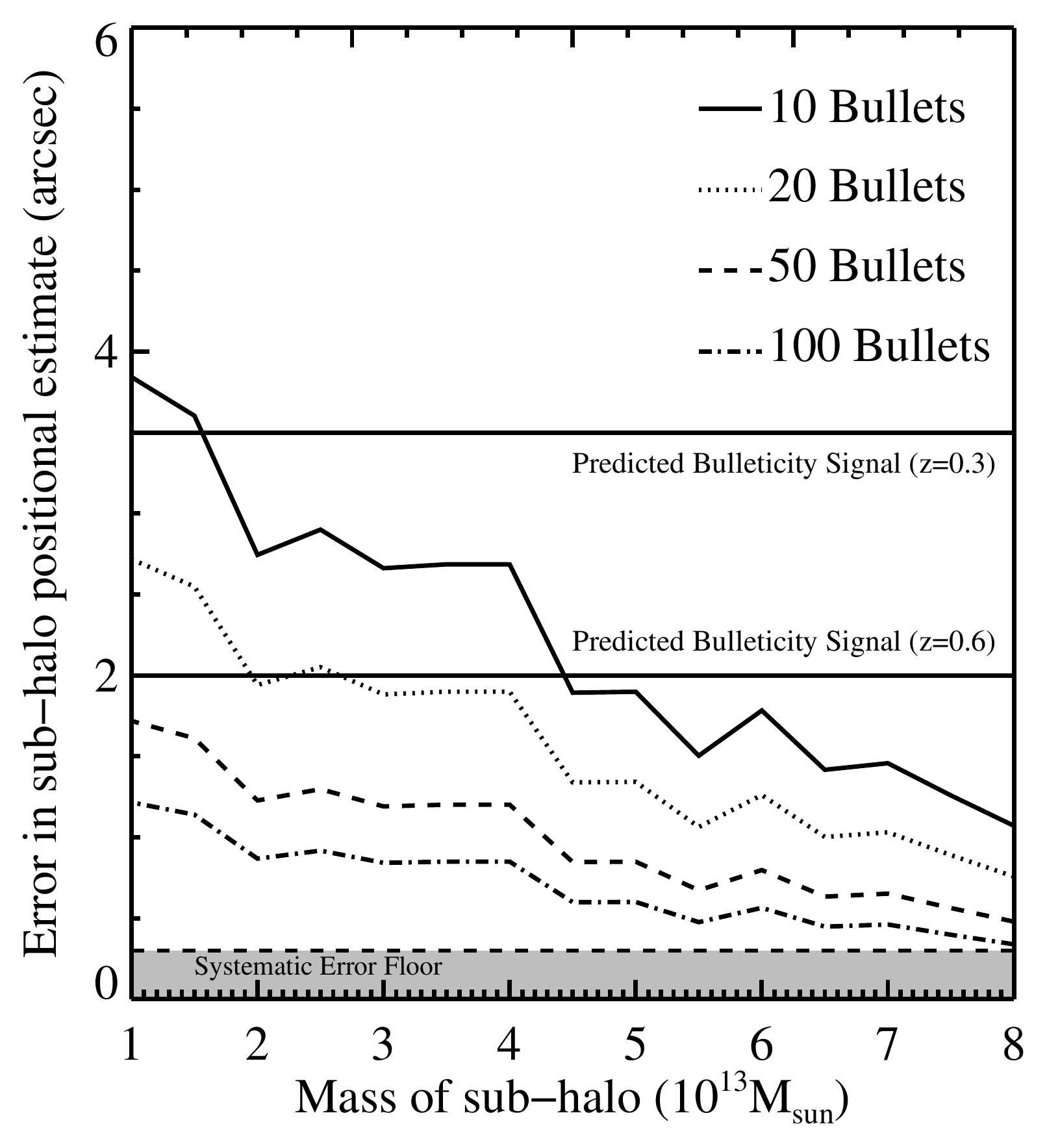}
 \caption{\label{fig:signal}
The error in the mean position for various sample sizes. In order to detect an overall offset between baryonic and dark matter the error in the mean of the sample size needs to be less than the expected signal. For clusters at redshift $z=0.6$, the bulleticity is $\sim2\arcsec$, and for a redshift $z=0.3$ this increases to $3.5\arcsec$. A sample of $\sim50$ offsets should yield a significant detection of bulleticity, and {\tt Lenstool} can measure these offsets with subdominant systematic bias.}
\end{figure}

\citet{2012MNRAS.419.3547D} find that cosmic shear can offset the position of weak lensing peaks of order $5$kpc ($0.7\arcsec$ at a $z=0.6$). We expect such a contaminant to average out to zero but have a contribution to the overall error and precision.
We therefore add this contaminant in quadrature to the error bars given in and calculate for a given size of clusters of a given mass with a given mass sub-halo for a cluster at redshift $z=0.6$. Figure \ref{fig:signal} shows between $\sim20-50$ measured offsets are required in order to have statistically significant detection. In this scenario there is also shape measurement bias and a source galaxy redshift distribution. 

Since typical clusters each have conservatively 1 in-falling group of galaxies containing $\sim10\%$ of their mass, this suggests between $20\sim50$ clusters are needed.
This is feasible within the current HST and Chandra archive. 
Furthermore, any strong lensing detections would tighten the constraints on the mass and concentration of the main halo can be more tightly constrained, and will lead to a better measurement of the offset. It is would be trivial to include a strong lensing model in to {\tt Lenstool}, however relative weighting of the constraints provided is an issue which will be addressed in future work.

Although basing predicted sample sizes on controlled environments such as those studied here, the results give us optimism to carry out the measurement in real data.

In the case of definite peaks and well defined profiles we expect a sample of between $20\sim50$ to be sufficient in order to measure a significant offset. We use this sample size as a confirmation that one should be able to make a detection using the current Hubble archive. We expect the true sample size to be larger than this and using hydrodynamical simulations, a more accurate sample size can be determined. Such tests are beyond the scope of this paper and will be carried out in conjunction with the data.

\section{Discussion}\label{sec:disc}

Through carefully controlled experiments, it was found that the likelihood surfaces for the reconstructed sub-halo positions are symmetric around the true value in the regime of infinite signal to noise.
In the presence of trivial noise contaminants the estimated positions are also not biased, however adding shape measurement bias seems to introduce a small bias of order $\sim0.3\arcsec$ in the sub-halo.

In order to better constrain our errors and any bias in position we averaged each mass scale over each single sub-halo configuration, and dual halo configuration.
It was found that the positional bias in all cases is independent of mass and configuration.
For a single halo configuration the only cause of bias was due to imperfect shape measurement. The observed offset of  $0.27\pm0.14$ is well within our tolerated level.
This bias was seen throughout the simulation using NFW profiles, including the dual sub halo configurations.
In the case of an SIS profile, we found that the positional estimate no longer observed a bias, which was due to the 
peaky nature of the central core, however the errors become unacceptably large below $1.5\times10^{13}M_\odot$.


In all of the simulations we do not find strong evidence for a positional bias of greater than $0.5\arcsec$. Importantly, this performance is sufficient to enable a detection of the theoretically expected $\sim2.0\arcsec$  ($\sim3.5\arcsec$) offset between substructure's dark matter and gas as it falls into massive clusters at a redshift of 0.6 (0.3) \citep{2011MNRAS.413.1709M}.

Initial work using the HST archive will aim to measure the average displacement between dark matter and both gas and stars, that could provide evidence for self interacting dark matter.
This displacement can then be calibrated with simulations of interacting dark matter to estimate its cross section.

With a sample of clusters already available in the HST archive, averaging the measured offset of many pieces of substructure will provide sufficiently accurate dark matter centroiding to detect an offset. 
Future space missions (e.g. Euclid$^1$\footnotetext[1]{http://www.euclid-ec.org} \citep{2011arXiv1110.3193L}) will increase the sample of available clusters by many orders of magnitude. However fully exploiting such data would require improved mass reconstruction techniques 

Furthermore, in the quasi-weak lensing regime considered here, flexion, the third derivative of the lensing potential, becomes important. 
As the gradient of the tidal field it is more sensitive to small scale structure, similar to that investigated here\citep{2006MNRAS.365..414B}.
The positional information in this higher order process could therefore provide significantly improved offset measurements. 
 Unfortunately, flexion remains extremely difficult to measure and it fundamental properties such as the intrinsic flexion distribution and accuracy requirements are still yet to be determined \citep{2012MNRAS.419.2215V}.
Future algorithms could potentially exploit this extra information however this is currently not possible.

\section{Conclusions}\label{sec:conc}

Measuring the separation of dark matter and baryonic gas in groups of infalling galaxies requires accurate astrometry. In the introduction we proposed five primary questions,
\begin{itemize}
\item To what precision can we measure the offset in a single infalling substructure? 
\item Can we identify a point estimator whose simple combination over a sample will provide a measurement of sub-halo position with minimal bias?
\item What are the dominant sources of residual bias in this estimate?
\item How large a sample of observed clusters are we likely to need to be able to detect an offset between dark matter and baryonic as predicted by MKN?
\item What further investigation might be needed to prove the utility of this technique for probing DM interaction cross-sections?
\end{itemize}
In reference to these, we find that
\begin{itemize}
\item  The public {\tt Lenstool} software can measure the position of individual $1.5 \times10^{13}M_\odot$ peaks with  $\sim0.3\arcsec$ systematic bias, as long as they are at least $\sim30\arcsec$ from the cluster centre. Any sub-halos detected above this threshold will be real and only biased to $\sim0.3\arcsec$.
\item  The maximum likelihood value of the 2 dimensional position likelihood surface is found to be the best point source estimator, being negligibly biased in the noise free case compared to the mean value estimator.
\item The dominant source of bias is caused by a preferred direction to the shape of galaxies introduced by an biased shape measurement algorithm. 
\item Since typical clusters each have on average 1 infalling groups of galaxies containing $\sim10\%$ of their mass, between $20-50$ clusters are needed to detect an offset between dark and baryonic matter.
\item The method will need to be tested on full hydrodynamical simulations (containing a more complex distribution of mass) in parallel with real data to show that the displacement obtained from data is reliable.
\end{itemize}
This work gives us confidence to pursue offsets as a technique in the measurement of the DM cross section.

\section*{Acknowledgments} 

The authors are pleased to thank Massimo Viola, Catherine Heymans and Jim Dunlop for useful conversations and advice.
DH is supported by an STFC studentship. RM, TK and PJM are supported by Royal Society URFs. EJ is supported by CNES.



\appendix

\bsp
\label{lastpage}


\begin{thebibliography}{99} 
\bibitem[\protect\citeauthoryear{Angloher et al.}{2011}]{2011arXiv1109.0702A} Angloher G., et al., 2011, arXiv, arXiv:1109.0702 
\bibitem[\protect\citeauthoryear{Baer \& Tata}{2009}]{2009patl.book..179B} Baer H., Tata X., 2009, Physics at the Large Hadron Collider, ISBN 978-81-8489-215-4. Springer India, 179 
\bibitem[\protect\citeauthoryear{Bacon et al.}{2006}]{2006MNRAS.365..414B} Bacon D.~J., Goldberg D.~M., Rowe B.~T.~P., Taylor A.~N., 2006, MNRAS, 365, 414 
\bibitem[\protect\citeauthoryear{Bartelmann \& Schneider}{2001}]{2001PhR...340..291B} Bartelmann M., Schneider P., 2001, Physics Reports, 340, 291 
\bibitem[\protect\citeauthoryear{Battaner \& Florido}{2000}]{2000FCPh...21....1B} Battaner E., Florido E., 2000, Fundamentals of Cosmic Physics, 21, 1 
\bibitem[\protect\citeauthoryear{Bernabei et al.}{2010}]{2010EPJC...67...39B} Bernabei R., et al., 2010, The European Physical Journal C, 67, 39 
\bibitem[\protect\citeauthoryear{Bertone, Hooper \& Silk}{2005}]{2005PhR...405..279B} Bertone G., Hooper D., Silk J., 2005,  Physics Reports, 405, 279 
\bibitem[\protect\citeauthoryear{Brada{\v c} et al.}{2006}]{2006ApJ...652..937B} Brada{\v c} M., et al., 2006, ApJ, 652, 937 
\bibitem[\protect\citeauthoryear{Brada{\v c} et al.}{2008}]{2008ApJ...687..959B} Brada{\v c} M., Allen S.~W., Treu T., Ebeling H., Massey R., Morris R.~G., von der Linden A., Applegate D., 2008, ApJ, 687, 959
\bibitem[\protect\citeauthoryear{Brada{\v c} et al.}{2005}]{2005A&A...437...39B} Brada{\v c} M., Schneider P., Lombardi M., Erben T., 2005, A\&A, 437, 39
\bibitem[\protect\citeauthoryear{Bridle et al.}{2010}]{2010MNRAS.405.2044B} Bridle S., et al., 2010, MNRAS, 405, 2044 
\bibitem[\protect\citeauthoryear{Burgos et al.}{2009}]{2009APh....31..261B} Burgos S., et al., 2009, APh, 31, 261
\bibitem[\protect\citeauthoryear{Cacciato et al.}{2006}]{2006A&A...458..349C} Cacciato M., Bartelmann M., Meneghetti M., Moscardini L., 2006, A\&A, 458, 349 
\bibitem[\protect\citeauthoryear{Cholis et al.}{2009}]{2009PhRvD..80l3511C} Cholis I., Goodenough L., Hooper D., Simet M., Weiner N., 2009, Physical Review D, 80, 123511 
\bibitem[\protect\citeauthoryear{Cohn}{2012}]{2012MNRAS.419.1017C} Cohn J.~D., 2012, MNRAS, 419, 1017
\bibitem[\protect\citeauthoryear{Clowe, Gonzalez \& Markevitch}{2004}]{2004ApJ...604..596C} Clowe D., Gonzalez A., Markevitch M., 2004, ApJ, 604, 596 
\bibitem[\protect\citeauthoryear{Clowe et al.}{2006}]{2006ApJ...648L.109C} Clowe D., Brada{\v c} M., Gonzalez A.~H., Markevitch M., Randall S.~W., Jones C., Zaritsky D., 2006, ApJ, 648, L109 
\bibitem[\protect\citeauthoryear{Diego et al.}{2007}]{2007MNRAS.375..958D} Diego J.~M., Tegmark M., Protopapas P., Sandvik H.~B., 2007, MNRAS, 375, 958 
\bibitem[\protect\citeauthoryear{Dietrich et al.}{2012}]{2012MNRAS.419.3547D} Dietrich J.~P., B{\"o}hnert A., Lombardi M., Hilbert S., Hartlap J., 2012, MNRAS, 419, 3547
\bibitem[\protect\citeauthoryear{Duffy et al.}{2010}]{2010MNRAS.405.2161D} Duffy A.~R., Schaye J., Kay S.~T., Dalla Vecchia C., Battye R.~A., Booth C.~M., 2010, MNRAS, 405, 2161
\bibitem[\protect\citeauthoryear{Finoguenov et al.}{2007}]{2007ApJS..172..182F} Finoguenov A., et al., 2007, ApJS, 172, 182 
\bibitem[\protect\citeauthoryear{Heymans et al.}{2006}]{2006MNRAS.368.1323H} Heymans C., et al., 2006, MNRAS, 368, 1323 
\bibitem[\protect\citeauthoryear{Hoekstra}{2001}]{2001A&A...370..743H} Hoekstra H., 2001, A\&A, 370, 743 
\bibitem[\protect\citeauthoryear{Hoekstra \& Jain}{2008}]{2008ARNPS..58...99H} Hoekstra H., Jain B., 2008, Annual Review of Nuclear and Particle Systems, 58, 99
\bibitem[\protect\citeauthoryear{Hoekstra et al.}{2011}]{2011ApJ...726...48H} Hoekstra H., Donahue M., Conselice C.~J., McNamara B.~R., Voit G.~M., 2011, ApJ, 726, 48
\bibitem[\protect\citeauthoryear{Hooper \& Goodenough}{2011}]{2011PhLB..697..412H} Hooper D., Goodenough L., 2011, Physics Letters B, 697, 412 
\bibitem[\protect\citeauthoryear{Jetzer et al.}{2002}]{2002astro.ph..1421J} Jetzer P., Koch P., Piffaretti R., Puy D., Schindler S., 2002, astro, arXiv:astro-ph/0201421
\bibitem[\protect\citeauthoryear{Jullo et al.}{2007}]{2007NJPh....9..447J} Jullo E., Kneib J.-P., Limousin M., El{\'{\i}}asd{\'o}ttir {\'A}., Marshall P.~J., Verdugo T., 2007, New Journal of Physics, 9, 447
\bibitem[\protect\citeauthoryear{Kitching et al.}{2012a}]{G10results} Kitching T.\ et al., 2012, MNRAS in press, arXiv:1202.5254
\bibitem[\protect\citeauthoryear{Kitching et al.}{2012b}]{MDM} Kitching T.\ et al., 2012, MNRAS in press, arXiv:1204.4096
\bibitem[\protect\citeauthoryear{Lasky \& Fluke}{2009}]{2009MNRAS.396.2257L} Lasky P.~D., Fluke C.~J., 2009, MNRAS, 396, 2257 
\bibitem[\protect\citeauthoryear{Laureijs et al.}{2011}]{2011arXiv1110.3193L} Laureijs R., et al., 2011, arXiv, arXiv:1110.3193 
\bibitem[\protect\citeauthoryear{Leauthaud et al.}{2007}]{2007ApJS..172..219L} Leauthaud A., et al., 2007, ApJS, 172, 219 
\bibitem[\protect\citeauthoryear{Leauthaud et al.}{2010}]{2010ApJ...709...97L} Leauthaud A., et al., 2010, ApJ, 709, 97 
\bibitem[\protect\citeauthoryear{Macci{\`o}, Dutton \& van den Bosch}{2008}]{2008MNRAS.391.1940M} Macci{\`o} A.~V., Dutton A.~A., van den Bosch F.~C., 2008, MNRAS, 391, 1940
\bibitem[\protect\citeauthoryear{Mahdavi et al.}{2007}]{2007ApJ...668..806M} Mahdavi A., Hoekstra H., Babul A., Balam D.~D., Capak P.~L., 2007, ApJ, 668, 806 
\bibitem[\protect\citeauthoryear{Massey et al}{2007}]{2007MNRAS.376...13M} Massey R.\ et al., 2007, MNRAS, 376, 13 
\bibitem[\protect\citeauthoryear{Massey et al.}{2007}]{2007Natur.445..286M} Massey R., et al., 2007, Nature, 445, 286
\bibitem[\protect\citeauthoryear{Massey \& Goldberg}{2008}]{2008ApJ...673L.111M} Massey R., Goldberg D.~M., 2008, ApJ, 673, L111 
\bibitem[\protect\citeauthoryear{Massey, Kitching \& Richard}{2010}]{2010RPPh...73h6901M} Massey R., Kitching T., Richard J., 2010, Reports on Progress in Physics, 73, 086901
\bibitem[\protect\citeauthoryear{Massey, Kitching \& Nagai}{2011}]{2011MNRAS.413.1709M} Massey R., Kitching T., Nagai D., 2011, MNRAS, 413, 1709 
\bibitem[\protect\citeauthoryear{Markevitch et al.}{2004}]{2004ApJ...606..819M} Markevitch M., Gonzalez A.~H., Clowe D., Vikhlinin A., Forman W., Jones C., Murray S., Tucker W., 2004, ApJ, 606, 819 
\bibitem[\protect\citeauthoryear{Melchior \& Viola}{2012}]{2012MNRAS.tmp.3383M} Melchior P., Viola M., 2012, MNRAS, 3383 
\bibitem[\protect\citeauthoryear{Merten et al.}{2009}]{2009A&A...500..681M} Merten J., Cacciato M., Meneghetti M., Mignone C., Bartelmann M., 2009, A\&A, 490, 681 
\bibitem[\protect\citeauthoryear{Merten et al.}{2011}]{2011MNRAS.417..333M} Merten J.\ et al., 2011, MNRAS, 417, 333 
\bibitem[\protect\citeauthoryear{Mitsou}{2011}]{2011JPhCS.335a2003M} Mitsou V.~A., 2011, Journal of Physics: Conference Series, 335, 012003
\bibitem[\protect\citeauthoryear{Nagai et al.}{2007a}]{nagai07a} Nagai D., Vikhlinin A.\ \& Kravtsov A., 2007a, ApJ 655, 98
\bibitem[\protect\citeauthoryear{Nagai et al.}{2007b}]{nagai07b} Nagai D., Kravtsov A.\ \& Vikhlinin A., 2007b, ApJ 668, 1
\bibitem[\protect\citeauthoryear{Navarro, Frenk \& White}{1996}]{1996ApJ...462..563N} Navarro J.~F., Frenk C.~S., White S.~D.~M., 1996, ApJ, 462, 563 
\bibitem[\protect\citeauthoryear{Peter et al.}{2012}]{2012arXiv1208.3026P} Peter A.~H.~G., Rocha M., Bullock J.~S., Kaplinghat M., 2012, arXiv, arXiv:1208.3026 
\bibitem[\protect\citeauthoryear{Powell, Kay \& Babul}{2009}]{2009MNRAS.400..705P} Powell L.~C., Kay S.~T., Babul A., 2009, MNRAS, 400, 705
\bibitem[\protect\citeauthoryear{Randall et al.}{2009}]{2009ApJ...700.1404R} Randall S.~W., Jones C., Markevitch M., Blanton E.~L., Nulsen P.~E.~J., Forman W.~R., 2009,ApJ, 700, 1404
\bibitem[\protect\citeauthoryear{Refregier}{2003}]{2003ARA&A..41..645R}Refregier A., 2003, ARA\&A, 41, 645
\bibitem[\protect\citeauthoryear{Rhodes et al.}{2007}]{2007ApJS..172..203R}Rhodes J.\ et al., 2007, ApJS 172, 203
\bibitem[\protect\citeauthoryear{Rocha et al.}{2012}]{2012arXiv1208.3025R} Rocha M., Peter A.~H.~G., Bullock J.~S., Kaplinghat M., Garrison-Kimmel S., Onorbe J., Moustakas L.~A., 2012, arXiv, arXiv:1208.3025
\bibitem[\protect\citeauthoryear{Rubin, Ford \& Thonnard}{1980}]{1980ApJ...238..471R} Rubin V.~C., Ford W.~K.~J., .~Thonnard N., 1980, ApJ, 238, 471 
\bibitem[\protect\citeauthoryear{Schmidt et al.}{2012}]{2012ApJ...744L..22S} Schmidt F., Leauthaud A., Massey R., 
Rhodes J., George M.~R., Koekemoer A.~M., Finoguenov A., Tanaka M., 2012, ApJ, 744, L22 
\bibitem[\protect\citeauthoryear{Scoville et al.}{2007}]{2007ApJS..172....1S} Scoville N., et al., 2007, ApJS, 172, 1 
\bibitem[\protect\citeauthoryear{Shan, Qin \& Zhao}{2010}]{2010MNRAS.408.1277S} Shan H.~Y., Qin B., Zhao H.~S., 2010, MNRAS, 408, 1277
\bibitem[\protect\citeauthoryear{Spinelli et al.}{2012}]{2012MNRAS.420.1384S} Spinelli P.~F., Seitz S., Lerchster M., Brimioulle F., Finoguenov A., 2012, MNRAS, 420, 1384 
\bibitem[\protect\citeauthoryear{Taylor et al.}{2007}]{2007MNRAS.374.1377T} Taylor A.~N., Kitching T.~D., Bacon D.~J., Heavens A.~F., 2007, MNRAS, 374, 1377 
\bibitem[\protect\citeauthoryear{Viola, Melchior, \& Bartelmann}{2012}]{2012MNRAS.419.2215V} Viola M., Melchior P., Bartelmann M., 2012, MNRAS, 419, 2215 
\bibitem[\protect\citeauthoryear{Williams \& Saha}{2011}]{2011MNRAS.415..448W} Williams L.~L.~R., Saha P., 2011, MNRAS, 415, 448
\bibitem[\protect\citeauthoryear{Wright \& Brainerd}{2000}]{2000ApJ...534...34W} Wright C.~O., Brainerd T.~G., 2000, ApJ, 534, 34 
\bibitem[\protect\citeauthoryear{Yoshida et al.}{2000}]{2000ApJ...544L..87Y} Yoshida N., Springel V., White S.~D.~M., Tormen G., 2000, ApJ, 544, L87 
\bibitem[\protect\citeauthoryear{Zwicky}{1933}]{1933AcHPh...6..110Z} Zwicky F., 1933, Helvetica Physica Acta, 6, 110
\end{thebibliography}
\end{document}